\documentclass[a4paper,UKenglish,cleveref, autoref, thm-restate]{lipics-v2021}

\hideLIPIcs  %

\usepackage{booktabs}
\usepackage{seqsplit} %
\usepackage{enumitem}
\newlist{romanenumeratea}{enumerate}{1}
\setlist[romanenumeratea]{label=\roman*., topsep=2pt, itemsep=1pt, parsep=1pt}
\usepackage{graphicx}
\usepackage[utf8]{inputenc}
\usepackage{longtable}
\usepackage{array}
\usepackage[T1]{fontenc}
\usepackage{caption}
\usepackage{subcaption}
\usepackage{listings}
\usepackage{algorithm}
\usepackage{algpseudocode}
\usepackage{xcolor}
\usepackage{hyperref}
\usepackage{tabularx}   %
\usepackage{siunitx} 
\usepackage{booktabs}
\usepackage{makecell}
\usepackage{graphicx}
\usepackage{array}
\colorlet{punct}{red!60!black}
\definecolor{background}{HTML}{EEEEEE}
\definecolor{delim}{RGB}{20,105,176}
\colorlet{numb}{magenta!60!black}

\lstdefinelanguage{json}{
    basicstyle=\normalfont\ttfamily,
    numbers=left,
    numberstyle=\scriptsize,
    stepnumber=1,
    numbersep=8pt,
    showstringspaces=false,
    breaklines=true,
    frame=lines,
    backgroundcolor=\color{background},
    literate=
     *{:}{{{\color{punct}{:}}}}{1}
      {,}{{{\color{punct}{,}}}}{1}
      {\{}{{{\color{delim}{\{}}}}{1}
      {\}}{{{\color{delim}{\}}}}}{1}
      {[}{{{\color{delim}{[}}}}{1}
      {]}{{{\color{delim}{]}}}}{1},
}
\lstdefinestyle{jsoncompact}{
  language      = json,
  numbers       = none,  
  captionpos = b,   
  basicstyle    = \ttfamily\small,
  showstringspaces = false,
  columns       = fullflexible,
  aboveskip     = 2pt,           %
  belowskip     = 0pt,
}

\title{Optimistic MEV in Ethereum Layer 2s:\\ Why Blockspace Is Always in Demand
}

\titlerunning{Optimistic MEV in Ethereum Layer 2s: Why Blockspace Is Always in Demand} 

\author{Ozan Solmaz}{ETH Zurich}{osolmaz@student.ethz.ch}{}{}
\author{Lioba Heimbach}{ETH Zurich}{hlioba@ethz.ch}{https://orcid.org/0000-0002-8258-1712}{}
\author{Yann Vonlanthen}{ETH Zurich}{yvonlanthen@ethz.ch}{https://orcid.org/0000-0001-5736-8197}{}
\author{Roger Wattenhofer}{ETH Zurich}{wattenhofer@ethz.ch}{https://orcid.org/0000-0002-6339-3134}{}

\authorrunning{O. Solmaz, L. Heimbach, Y. Vonlanthen and R. Wattenhofer} 

\Copyright{CC-BY} %



\category{} %

\acknowledgements{We thank Burak Öz, Vabuk Pahari and the anonymous AFT reviewers for their helpful feedback. Our gratitude extends to Patrick Züst for his exploratory work in his Master thesis supervised by the last three authors and Quentin Kniep, which helped shape this work. }%

\nolinenumbers %

\EventEditors{Zeta Avarikioti and Nicolas Christin}
\EventNoEds{2}
\EventLongTitle{7th Conference on Advances in Financial Technologies
(AFT 2025)}
\EventShortTitle{AFT 2025}
\EventAcronym{AFT}
\EventYear{2025}
\EventDate{October 8--10, 2025}
\EventLocation{Pittsburgh, PA, USA}
\EventLogo{}
\SeriesVolume{354}

\begin{document}

\maketitle

\begin{abstract}

Layer 2 rollups are rapidly absorbing DeFi activity, securing over \$40 billion and accounting for nearly half of Ethereum’s DEX volume by Q1 2025, yet their MEV dynamics remain understudied. We address this gap by defining and quantifying optimistic MEV, a form of speculative, on-chain MEV whose detection and execution logic reside largely on-chain in smart contracts. As a result of their speculative nature and lack of off-chain opportunity verification, optimistic MEV transactions frequently decide not to execute any trades.

In this work, we focus on cyclic arbitrage, which we find is predominantly executed as optimistic MEV on Layer 2s. Using our multi-stage identification pipeline on Arbitrum, Base, and Optimism, we show that in Q1 2025, transactions from cyclic arbitrage contracts account for over 50\% of on-chain gas on Base and Optimism and 7\% on Arbitrum, driven mainly by ``interaction'' probes (on-chain computations searching for arbitrage). This speculative probing indicates that cyclic arbitrage on Layer 2s is predominantly executed as optimistic MEV and contributes to generally keeping blocks on Base and Optimism persistently full. Despite consuming over half of on-chain gas, these optimistic MEV transactions pay less than one quarter of total gas fees. Cross-network comparison reveals divergent success rates, differing patterns of code reuse, and sensitivity to varying sequencer ordering and block production times. Finally, OLS regressions link optimistic MEV trade count to ETH volatility, retail trading activity, and DEX aggregator usage. Together, these findings show that optimistic MEV has become a major source of persistent spam-like transaction activity on Layer 2s, dominating blockspace with low-value probes and reshaping the composition of on-chain activity.
\end{abstract}

\section{Introduction}

Ethereum, a decentralized and programmable blockchain, features robust smart contract functionality that enables trust-minimized applications and value transfer without reliance on traditional intermediaries. This programmability has cultivated a dynamic ecosystem of decentralized applications, particularly within the sphere of \textit{decentralized finance (DeFi)}, which includes \textit{decentralized exchanges (DEXes)} and \textit{lending protocols}. As DeFi and other on-chain activities have grown in scale and value, they have contributed to significant network congestion on Ethereum Layer 1, driving transaction fees to levels that render smaller-scale operations economically unviable.

To address these critical scaling limitations, the Ethereum community has strategically embraced a rollup-centric development roadmap~\cite{buterin2024,polynya2025,heimbach2025earlydaysethereumblob}.
Layer 2 rollups scale Ethereum by moving transaction execution off-chain and periodically anchoring summarized results on-chain.
This design preserves security while enabling higher throughput and lower costs~\cite{growthepie2025,l2beat2025}. Adoption has been significant: as of April 2025, Layer 2 networks secure around \$40~B in assets and account for a growing share of on-chain transaction volume, highlighting their central role in the Ethereum ecosystem.

\begin{figure}[t]\vspace{-10pt}
    \centering
    \includegraphics[width=\linewidth]{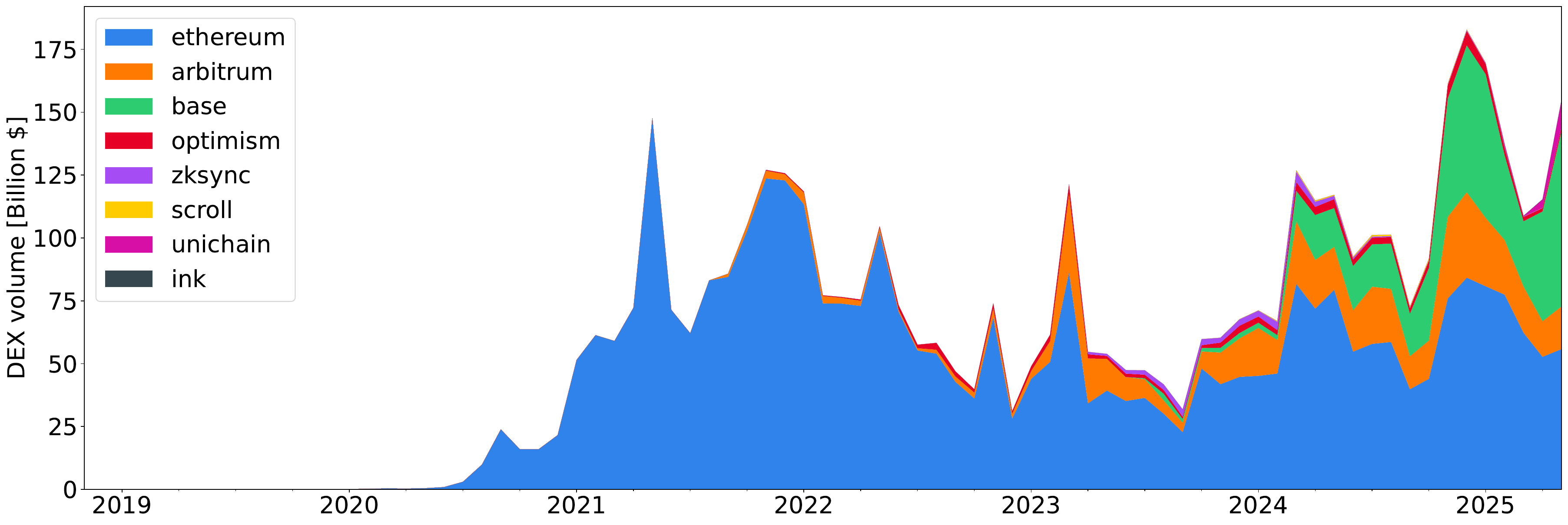}
    \caption{Monthly DEX transaction volume by network from November 2018 to May 2025. The plot highlights the significant rise of Layer 2 networks --- such as Arbitrum, Base, Optimism, and others --- in total DEX activity, with Ethereum Layer 1 gradually representing a smaller share of the overall volume.}
    \label{fig:volume}
\end{figure}

As illustrated in Figure~\ref{fig:volume}, a substantial share of DeFi activity now occurs on Layer 2 networks. To be precise, in the first quarter of 2025, roughly 47\% of DEX volume in the Ethereum ecosystem was happening on Layer 2 networks. This figure is up from 35\% in the first quarter of 2024. This migration of DeFi activity to Layer 2 networks has concomitantly fostered a distinct \textit{Maximal Extractable Value (MEV)} landscape.

The shift in the MEV landscape has been accelerated by the advent of Layer 2 rollups and upgrades such as EIP-4844 Proto-Danksharding~\cite{EIP4844Web2025,EIP4844GitHub2025}, which have significantly reduced data availability costs for Layer 2 networks and, in turn, lowered transaction fees on these networks. These lower costs have allowed for MEV extraction strategies that are less viable on the Ethereum mainnet. Beyond lower fees, Layer 2s introduce operational characteristics that further enable MEV strategies previously infeasible on Layer 1. These include differing transaction ordering policies implemented by sequencers --- ranging from simple \textit{First-Come, First-Served (FCFS)} to auction-based mechanisms such as \textit{Priority Gas Auctions (PGA)} --- as well as variations in mempool privacy, and short inter-block times.

The confluence of these features reduces the financial risk and increases the execution uncertainty of MEV-seeking transactions. As a result, cyclic arbitrage bots can adopt an optimistic approach: they issue high-frequency, speculative transactions without knowing in advance whether an arbitrage opportunity exists, relying instead on rapid on-chain state reads and being sufficiently close in time to the opportunity-creating transaction to back-run it successfully. This differs from traditional cyclic arbitrage on Ethereum Layer 1, where generally bots precompute a guaranteed profit path and bundle all trades into a single cyclic transaction. On Layer 2, optimistic strategies involve repeatedly probing liquidity pools for small gains, despite not knowing if an opportunity exists, and accepting a high failure rate in exchange for marginal profits.

We refer to this class of strategies that rely on speculative execution, where MEV bots submit transactions dependent on on-chain computation to identify profitable opportunities, often resulting in failed attempts, as \textit{optimistic MEV}. To the best of our knowledge, this is the first work to formalize and systematically study optimistic MEV on Layer 2s.

\subparagraph*{Our contributions.} We summarize our main contributions below.
\begin{romanenumeratea}[topsep=2pt, itemsep=1pt, parsep=1pt, partopsep=0pt]
    \item We provide the first definition of \textit{optimistic MEV}, describing it as a class of MEV extraction strategies that rely heavily on on-chain logic for opportunity identification and speculative execution. 
    \item  We design and implement a multi-stage pipeline to identify and classify MEV transactions. This includes the construction of a cyclic arbitrage detector, an on-chain behavior classifier, and a revert/success labeling system to capture speculative dynamics.
    \item  We apply our methodology to present the first large-scale measurement of optimistic MEV on major Layer 2s, showing that cyclic arbitrage accounts for roughly 7\% of gas usage on Arbitrum, 51\% on Base, and 55\% on Optimism in Q1 2025 and is predominantly optimistic in nature. In particular, out analysis reveals that a large share of this gas is consumed by transactions that fail to execute profitable trades, supporting the hypothesis that these bots engage in speculative on-chain probing.
    \item  We compare optimistic MEV activity across Arbitrum, Base, and Optimism and find substantial differences in strategy execution patterns, success rates, and gas usage concentration. We attribute these to network-specific factors such as transaction ordering policies and inter-block times.
    \item  Using regression analysis, we examine how optimistic MEV activity correlates with market conditions and user trade behavior. Our findings suggest that volatility, trade volume, and aggregator usage significantly impact optimistic MEV prevalence.
    \item We discuss the broader implications of optimistic MEV for the MEV ecosystem and Layer 2 design, highlighting its economic impact and the challenges it poses for existing mitigation approaches.
\end{romanenumeratea}

\section{Related Work}
\subparagraph{Maximal Extractable Value.}
The study of MEV was commenced by Eskandari et al. \cite{eskandari2019sok} and Daian et al.~\cite{daian2019flashboys20frontrunning}, who first defined MEV and documented phenomena like front-running and PGAs on Ethereum. In subsequent work, Qin et al.~\cite{qin2021quantifyingblockchainextractablevalue} and Torres et al.~\cite{ferreira2021frontrunner}, quantified various forms of MEV on Ethereum mainnet, including sandwich attacks (i.e., a type of front-running attack on DEXes), liquidations of positions on lending protocols, and cyclic arbitrage, providing a baseline understanding of its scale and impact. A subsequent line of work analyzes various and evolving aspects of the MEV landscape on the Ethereum Layer 1~\cite{zhou2021high,zhou2021just,ferreira2021frontrunner,wang2022cyclicarbitragedecentralizedexchanges,paradigmdarkforest,rektdarkforest,samchepalhidden,samczsunescaping,misakawormhole,weintraub2022flash,heimbach2023ethereum,öz2024winsethereumblockbuilding,heimbach2024nonatomicarbitragedecentralizedfinance}. While these works have established a foundation for understanding MEV on Ethereum Layer 1, the Layer 2 landscape remains comparatively underexplored. Our work builds on this foundation by focusing on the unique characteristics of Layer 2s, providing a classification and analysis of MEV transactions in these emerging environments.

A closely related line of work examines MEV in alternative Layer 1 networks, which operate under distinct architectural and economic conditions. Öz et al.~\cite{öz2024playingmevgamefirstcomefirstserved} analyze MEV on FCFS blockchains such as Algorand, identifying latency optimization --- rather than fee bidding --- as the dominant extraction mechanism. Further, work done by Umbra Research~\cite{umbramevsolana} has provided valuable insights into the MEV on Solana. They point to the existence of optimistic MEV on Solana but provide no in-depth analysis of the phenomenon. In contrast, our work focuses on Layer 2 networks within the Ethereum ecosystem, where optimistic MEV is both prevalent and, to the best of our knowledge, has not been investigated in depth.

\subparagraph{Maximal Extractable Value in Layer 2 Networks.}
Recent efforts have begun to examine MEV on Layer 2 networks. Torres et al.~\cite{torres2024rollingshadowsanalyzingextraction} conducted a comparative analysis of MEV across Ethereum and major Layer 2s (Arbitrum, Optimism, zkSync). Theoretical work has also addressed cross-domain MEV, with Obadia et al.~\cite{obadia2021unitystrengthformalizationcrossdomain} formalizing MEV across domains such as different Layer 2s. Complementing this, Gogol et al.~\cite{gogol2024crossrollupmevnonatomicarbitrage} empirically analyzed non-atomic cross-rollup MEV and CEX-DEX arbitrage. Öz et al.~\cite{oz2025pandora} present the first systematic study of non-atomic cross-chain arbitrage strategies across nine blockchains, including several Layer 2s. Their work highlights the effects of these strategies on network congestion and the security implications of cross-chain MEV. Our work extends these Layer 2-focused studies by analyzing optimistic MEV, a form of MEV that involves on-chain detection of opportunities.

\subparagraph*{Decentralized Exchanges.}
Theoretical frameworks for routing on DEXes and profits of liquidity providers have provided important context for understanding MEV behavior. Angeris et al.~\cite{angeris2022optimalroutingconstantfunction} analyzed optimal routing and arbitrage in CFMMs. Milionis et al.~\cite{milionis2023automatedmarketmakingarbitrage,milionis2024automatedmarketmakinglossversusrebalancing} modeled loss-versus-rebalancing (LVR) under Poisson-distributed interblock times; and Nezlobin et al.~\cite{nezlobin2025lossversusrebalancingdeterministicgeneralizedblocktimes} extended this to deterministic block intervals, particularly relevant given the regular block production in many blockchain networks. Building on these theoretical foundations, our work provides empirical insights into gas usage, MEV bot behavior, and protocol-level design factors across Layer 2 networks, focusing on optimistic MEV: a speculative, high-frequency form of MEV.

\section{Optimistic MEV}

Next, we introduce optimistic MEV: a class of extraction strategies that (1) perform opportunity discovery and trade execution primarily through on-chain computation and (2) rely on speculative transaction submission without prior off-chain verification of profitability.

This behavior is most evident in \textit{optimistic cyclic arbitrage}, the focus of this work. The MEV types under consideration are sandwich attacks, cyclic arbitrage (arbitrage between DEXes on the same chain), non-atomic arbitrage (e.g., CEX–DEX arbitrage), and liquidations of positions on lending protocols. Of these, cyclic arbitrage offers a practically viable case where opportunities can be identified and executed entirely on-chain in a single transaction through frequent probing. While liquidations are in principle detectable on-chain, they occur only when loans become undercollateralized and are likely too infrequent to support high-frequency probing~\cite{qin2021quantifyingblockchainextractablevalue}. Sandwich attacks require two separate transactions, and non-atomic arbitrage depends on external price data from off-chain venues. For this reason, we focus our analysis of optimistic MEV on cyclic arbitrage, as this MEV strategy can be detected and executed fully on-chain.

Empirically, we observe this form of MEV most prominently on low-fee Layer 2 networks. Its defining signature is speculative execution: bots submit transactions without certainty that an arbitrage opportunity exists, using cheap blockspace to make such probing economically viable. In our analysis, these transactions frequently begin with top-level \texttt{STATICCALL} operations directed at DEX liquidity pools to read real-time reserves and prices. The contract then decides whether to execute a swap or terminate without trading based on these on-chain checks. This behavior, combined with the common use of minimal or repeated \texttt{calldata}, indicates that key parameters, i.e., the path of the arbitrage transaction and the swap amounts, are determined during execution rather than precomputed off-chain. The reliance on such runtime inspection also explains why optimistic MEV becomes infeasible when transaction fees are high: the preparatory on-chain calls consume significant gas. These observations motivate the following definition of optimistic MEV.

\begin{definition}[\textbf{Optimistic MEV}]
\emph{Optimistic MEV} denotes MEV strategies in which the existence and parameters of a profitable opportunity are determined within the transaction itself through on-chain computation at execution time, rather than being verified off-chain prior to submission. This leads to speculative transactions where profitability is only resolved during execution, resulting in a significant share of transactions that terminate without performing a trade.
\end{definition}

\subparagraph*{Case Study.}
To concretely illustrate the observed Layer 2 MEV patterns, particularly the prevalent use of initial \texttt{STATICCALL} operations for on-chain reconnaissance, we present a comparative case study of two transactions interacting with the same contract \href{https://basescan.org/address/0xF5fF765b0c1278E54281193d7019281e0e50A8C0}{\texttt{0xf5ff\dots a8c0}}\footnote{We marked this contract as one of the top gas consuming cyclic arbitrage contracts on Base, which we will discuss in more detail in the later sections.}: \href{https://basescan.org/tx/0x1d977d6867e2868b518a10803d64b414e428bd8e639d3c5054b2529cb55d18cb}{\texttt{0x1d97\dots 18cb}} (henceforth $Tx_A$) and \href{https://basescan.org/tx/0xb67825a6fa60e4bd9892076ead93c41f631460a53b8219036a5ace051f139bd7}{\texttt{0xb678\dots 9bd7}} (henceforth $Tx_B$) on Base. Both transactions were initiated with the same function selector, namely \texttt{0x00003748}, as evidenced by the identical first $4$ bytes in the calldata. This common entry point suggests a shared execution pathway, likely designed for conditional arbitrage based on real-time market conditions. Despite this identical initiation, the on-chain outcomes of $Tx_A$ and $Tx_B$ diverged significantly. $Tx_A$ resulted in no token swaps. In stark contrast, $Tx_B$ successfully executed a 2-swap cyclic arbitrage, cycling value through the path $WETH \to TYBG \to WETH$ and realizing a profit.
\vspace{3mm}

\begin{lstlisting}[style=jsoncompact,
  caption={Condensed initial call sequence from $Tx_A$, illustrating
           \texttt{STATICCALL} probes to DEX pools. Input \texttt{slot0()}
           retrieves key state variables from Uniswap V3-compatible pools.},
  label={lst:trace_example}]
{ 
"calls": [
    { "input": "slot0()", "to": "UNIV3 ETH-USDC Pool", "type": "STATICCALL" },
    { "input": "slot0()", "to": "UNIV3 PLAY-USDC Pool", "type": "STATICCALL" },
    { "input": "slot0()", "to": "CL ETH-PLAY Pool",  "type": "STATICCALL" }
    // ... further calls omitted for brevity
  ],
  "from": "0xdd57...a88",
  "to":   "0xf5ff...8c0",
  "type": "CALL" 
  }
\end{lstlisting}

A detailed analysis of the execution trace for $Tx_A$ reveals that the contract systematically probed multiple DEX liquidity pools via \texttt{STATICCALL} operations. These read-only calls occurred before any attempt to execute a trade, and the transaction subsequently terminated. This sequence of actions is strongly indicative of an on-chain reconnaissance strategy, where the contract assesses current pool states (e.g., token reserves, potential slippage, pool's fee structure, or tick data) to determine the viability of an arbitrage opportunity before committing to an execution path. A condensed representation of the initial call sequence within $Tx_A$ is provided in Example~\ref{lst:trace_example}.

\section{Data Classification}
To study optimistic MEV, we start by collecting relevant data (see Appendix~\ref{app:datacollection}) from the Ethereum Layer 1 as well as Arbitrum, Base, and Optimism (i.e., the biggest three Layer 2s in terms of DEX volume at the time of this writing). Then, we systematically identify cyclic arbitrage MEV activities on Layer 2 networks using a three-stage classification methodology. Recall that we focus on optimistic cyclic arbitrage in our work, as cyclic arbitrage opportunity detection can be fully resolved on-chain in a single transaction. In Section~\ref{sec:optimistic}, we will then show that cyclic arbitrage on Layer 2 networks predominantly operates optimistically.

The implementation of our cyclic arbitrage classification pipeline, as well as the classification output (the full list of cyclic arbitrage contracts) are made openly accessible~\cite{non-anonymous2025optimistic}.

\subsection{Cyclic Arbitrage Contract Detection}
\label{sec:classification:stage1}
Building on the cyclic arbitrage detection heuristic of Wang et al.~\cite{wang2022cyclicarbitragedecentralizedexchanges} and the toolkits behind EigenPhi and \texttt{libmev}~\cite{eigenphi,libmev}, we implement a cyclic arbitrage contract detection pipeline in Dune SQL, leveraging Dune Analytics' data tables and the Torino engine.
Inspired by the replay logic of \texttt{mev-inspect-py}~\cite{Flashbots2025MEVInspect} and \textit{Brontes}~\cite{SorellaLabs2025Brontes} we adapt our methodology to the Layer 2 circumstances.

We start by computing the set of transactions that, with high likelihood, perform cyclic arbitrage:

\begin{enumerate}[topsep=2pt, itemsep=1pt, parsep=1pt, partopsep=0pt]
  \item \textbf{Token Path \&  Balance Reconstruction.}  
        We consider all transactions containing swap events by querying the swap logs from \texttt{dex.trades}~\cite{dune2025trades}. Each transaction’s swap events are ordered to recover the exact token path and the initiator’s net token balance.
  \item \textbf{Router \& Aggregator Filter.}  
        Transactions that directly interact with labeled routers (captured through the metadata of the  \texttt{dex.addresses}~\cite{dune2025labels} table) or aggregators (captured through the \texttt{dex\_aggregator.trades}~\cite{duneaggt2025} table) are dropped.
  \item \textbf{Cyclic Arbitrage \& Profit Filter.}  
        A transaction is kept only 
        if it forms a sequence of at least two swaps, such that (i) the the token bought in the $j$-th swap is the same as the token sold in the $(j+1)$-th swap,  (ii) the sequence begins and ends with the same token and (iii) yields a strictly non-negative balance change in every token, with at least one positive gain.
\end{enumerate}

The first callee in any such profitable, cyclic arbitrage transaction is tagged as a \emph{candidate contract}, indicating it likely belongs to a cyclic arbitrage bot.
A more detailed and formal description of this classification process can be found in \cref{app:classification}.

Next, we inspect the candidate contracts once again to avoid misclassifications. The few misclassified contracts we find during this stage are mainly due to popular mislabeled routers or public utility contracts. The details of this step can be found in \cref{app:manaudit}. Finally, the remaining contracts form our final set of cyclic arbitrage contracts $\mathcal{C}_{\text{bot}}$.

\subsection{Transaction-Level Classification}

We classify each transaction in our study window (obtained through the respective table $\texttt{\{\{chain\}\}.transactions}$~\cite{dune2025transactions}) according to three dimensions. 

\subparagraph*{Transaction purpose (\textsc{cyclicArb} / \textsc{other}).}  
\label{cat1}
The transaction is marked \textsc{cyclicArb} when its first callee (the address in the \texttt{to} field) is a member of the validated set of cyclic arbitrage contracts \(\mathcal{C}_{\text{bot}}\); otherwise, it is marked \textsc{other}.  This choice ties the label to the entity that originates the on-chain action rather than to the presence of any particular pattern inside the trace.

\subparagraph*{DEX involvement (\textsc{trade }/ \textsc{interaction} / \textsc{residual}).} 

We scan swaps in \texttt{dex.trades} and consider the full call trace (\texttt{\{\{chain\}\}.traces}).  
If the transaction itself emits at least one swap event (e.g.\ \texttt{Swap} in Uniswap V2/V3~\cite{uniswap} or \texttt{TokenExchange} in Curve~\cite{curvefi} as a result of function calls such as Uniswap V2's \texttt{swapExactTokensForTokens}, Uniswap V3 pool's \texttt{swap} method \cite{uniswap}, or a Curve pool's \texttt{exchange} function \cite{curvefi}), we label it \textsc{trade}.  
When no swap is emitted but the trace touches a recognized DEX pool contract, typically through read-only calls such as \texttt{getReserves} or \texttt{slot0}, the label assigned is \textsc{interaction}. (We use the table \texttt{dex.raw\_pools} to detect the existence of subcalls interacting with DEX-related contracts.) 
All remaining transactions, which never enter a DEX contract, are labelled \textsc{residual}.  
The three options are mutually exclusive and collectively exhaustive.

\subparagraph*{Execution outcome (\textsc{success} / \textsc{revert}).}  
Finally, this dimension records the final on-chain status of the transaction, based on its receipt. The transaction is marked as \textsc{success} when execution was successful (\texttt{status}=1) and  \textsc{revert} otherwise (\texttt{status}=0). A transaction classified as a \textsc{trade} cannot subsequently be classified as a \textsc{revert}, as only transactions that do not revert can perform trades.

These three categories give each transaction a concise profile, which we use to group transactions for the analysis that follows. For example, we write \textsc{cyclicArb} - \textsc{interaction} to mean the subset of transactions that invoke a cyclic arbitrage contract as the first callee and also probe a DEX contract.\footnote{One thing to note is that \textsc{cyclicArb} - \textsc{interaction} contains transactions from \textsc{cyclicArb} - \textsc{interaction} - \textsc{success} and \textsc{cyclicArb} - \textsc{interaction} - \textsc{revert}.} \textsc{other} - \textsc{revert} defines the set of all reverted transactions that have a callee outside of the set \(\mathcal{C}_{\text{bot}}\).  Comparing such sets of transactions over time and across roll-ups enables us to quantify the prevalence of optimistic MEV, success rates, and failure modes under different Layer 2 design choices, such as private mempools, sequencer ordering, low fees, and sub-second blocks.

\section{Optimistic MEV Landscape}\label{sec:optimistic}

We commence our analysis of optimistic MEV by providing a broad overview of the landscape. Our analysis focuses on the period from August 2023 to May 2025 and looks at three Layer 2 networks (i.e., Arbitrum, Base, and Optimism) in comparison to the Ethereum Layer 1.

\subsection{Execution Outcomes and Success Rates}
Recall that the data classification identified cyclic arbitrage contracts. The following analysis shows that on Layer 2 networks, these contracts operate predominantly in an optimistic manner, exhibiting the speculative on-chain behavior characteristic of optimistic MEV, in contrast to their counterparts on Ethereum Layer 1.

\begin{figure}[thb]\vspace{-10pt}
    \centering
    \includegraphics[width=\linewidth]{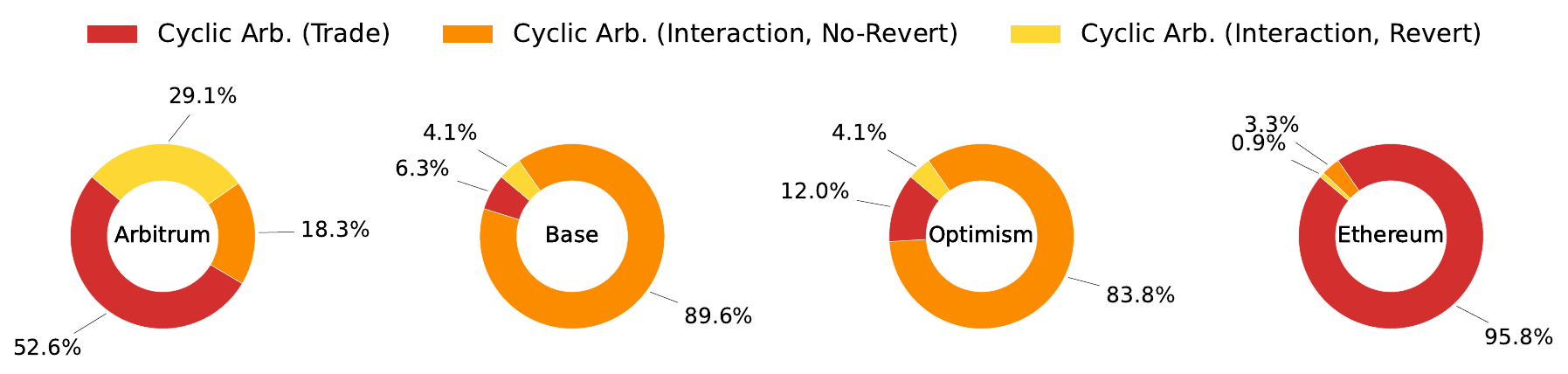}
    \caption{Distribution of cyclic arbitrage transaction outcomes across networks. Red indicates transactions from cyclic arbitrage contracts that execute trades, while the yellow and orange segments represent transactions that likely attempt arbitrage but do not execute any trade, either terminating without swaps (orange) or reverting entirely (yellow).}
    \label{fig:optimism}
\end{figure}

Figure~\ref{fig:optimism}  illustrates the relative share of transactions from identified cyclic arbitrage contracts that execute swaps (red) compared to those that only probe DEX pools and terminate without a trade (orange) or revert (yellow), considering only the subset of transactions that interact with a DEX. The difference between networks is pronounced. On Base and Optimism, only 6.3\% and 12\% of such transactions, respectively, result in a successful trade, whereas on Arbitrum the share rises to 52.6\% and on Ethereum Layer 1 to 95.8\%. This distribution highlights a fundamental contrast: on Layer 1, cyclic arbitrage transactions appear largely precomputed and succeed overwhelmingly, while on Layer 2s, and particularly on OP-stack rollups, they display a speculative character with most attempts ending without executing a trade. This pattern of profitability being resolved only at execution and the prevalence of transactions that terminate without trading is exactly what defines optimistic MEV. %

\begin{figure}[thb]
    \centering
    \includegraphics[width=\linewidth]{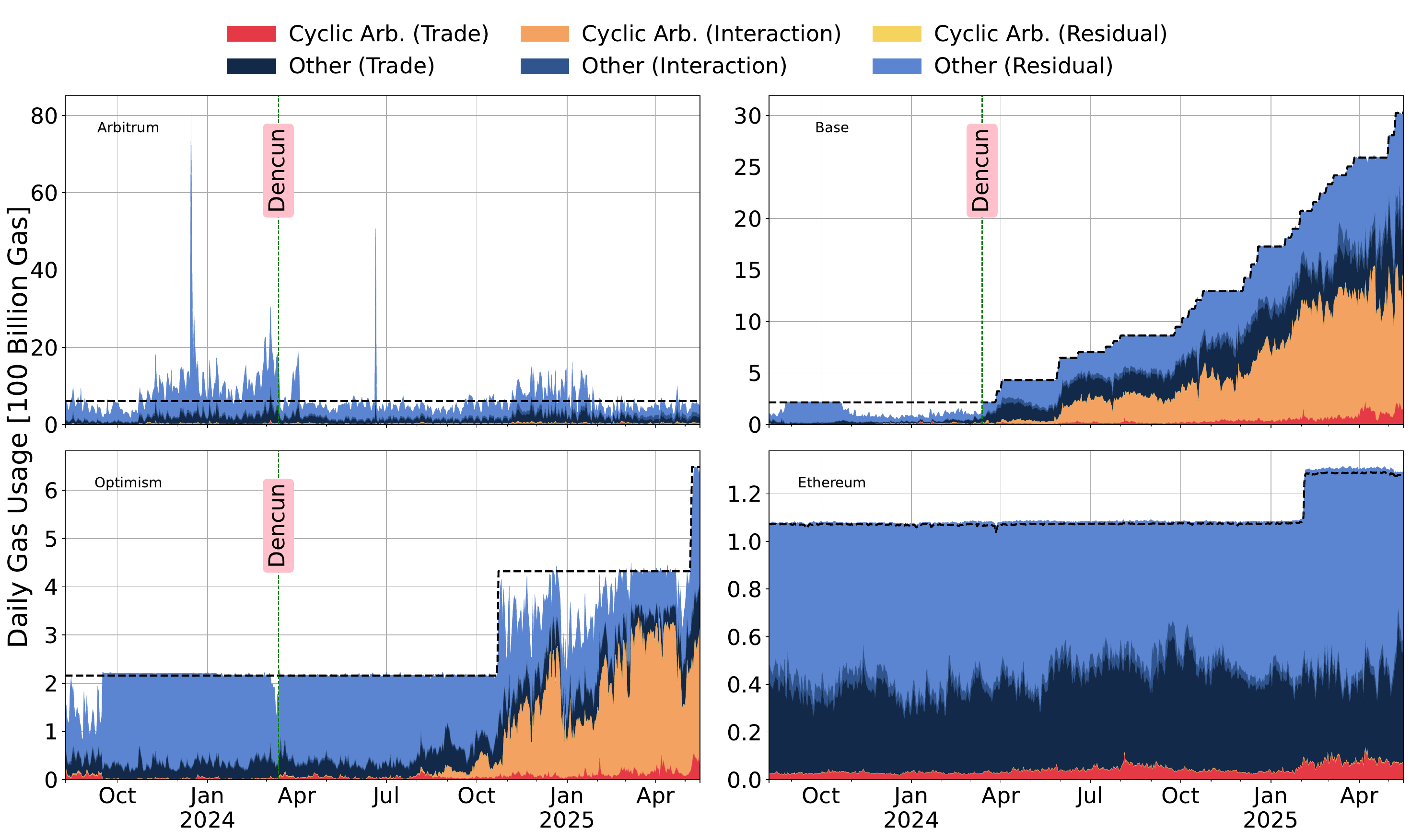}
    \caption{Daily gas usage by transaction category on Arbitrum, Base, Optimism, and Ethereum Layer 1. Each area shows gas consumed by cyclic arbitrage MEV bots (\textsc{cyclicArb}) and all other activity (\textsc{other}), subdivided into \textsc{trade} (transactions with executed swaps), \textsc{interaction} (on-chain probing without swaps), and \textsc{residual} segments (see Section~\ref{cat1}). The vertical green dotted line marks the Dencun upgrade (EIP-4844), while the black dashed line indicates each network’s target gas limit (``Gas Target'' on Base, Optimism, and Ethereum; ``Speed Limit'' on Arbitrum).}
    \label{fig:gas}
\end{figure}

\subsection{Prevalence Across Networks}

We explore these differences further in Figure~\ref{fig:gas}, showing the evolution of daily gas usage on Arbitrum, Base, and Optimism, broken down by activity from addresses performing \textsc{cyclicArb} MEV and all \textsc{other} activity. Each category is further subdivided into \textsc{trade} (transactions with executed token swaps), \textsc{interaction} (on-chain probing or contract calls without token transfers), and \textsc{residual} activity (see Section~\ref{cat1}). Recall that the \textsc{interaction} category includes both probing transactions that never execute a swap as well as reverted transactions. Notably, the share of gas consumed by \textsc{cyclicArb} MEV (yellow, orange, and red), which on Layer 2 networks manifests predominantly as optimistic MEV, is very large on Base and Optimism in particular. The share of gas attributed to \textsc{cyclicArb} - \textsc{interaction} transactions rises sharply over time, indicating a surge in speculative, non-token-transfer, DEX contract probing activity associated with optimistic MEV. On Base, this category becomes a dominant contributor to gas usage by early 2025, accounting for 48\% of total gas in the first quarter of the year. Overall, these optimistic MEV bots performing cyclic arbitrage are responsible for 51\% of Base’s gas usage during this period.  Further, gas usage from an optimistic MEV bot on Base exceeds the entire block capacity of Ethereum Layer 1 by an order of magnitude. Optimism exhibits a similar trend, though with a slightly delayed onset. By the first quarter of 2025, optimistic MEV bots performing cyclic arbitrage account for 55\% of gas usage on Optimism, with \textsc{cyclicArb} - \textsc{interaction} transactions specifically contributing 52\%. In contrast, on Arbitrum and Ethereum Layer 1, \textsc{cyclicArb} MEV consumes a much smaller fraction of available blockspace.

\begin{figure}[h]\vspace{-10pt}
    \centering
    \includegraphics[width=\linewidth]{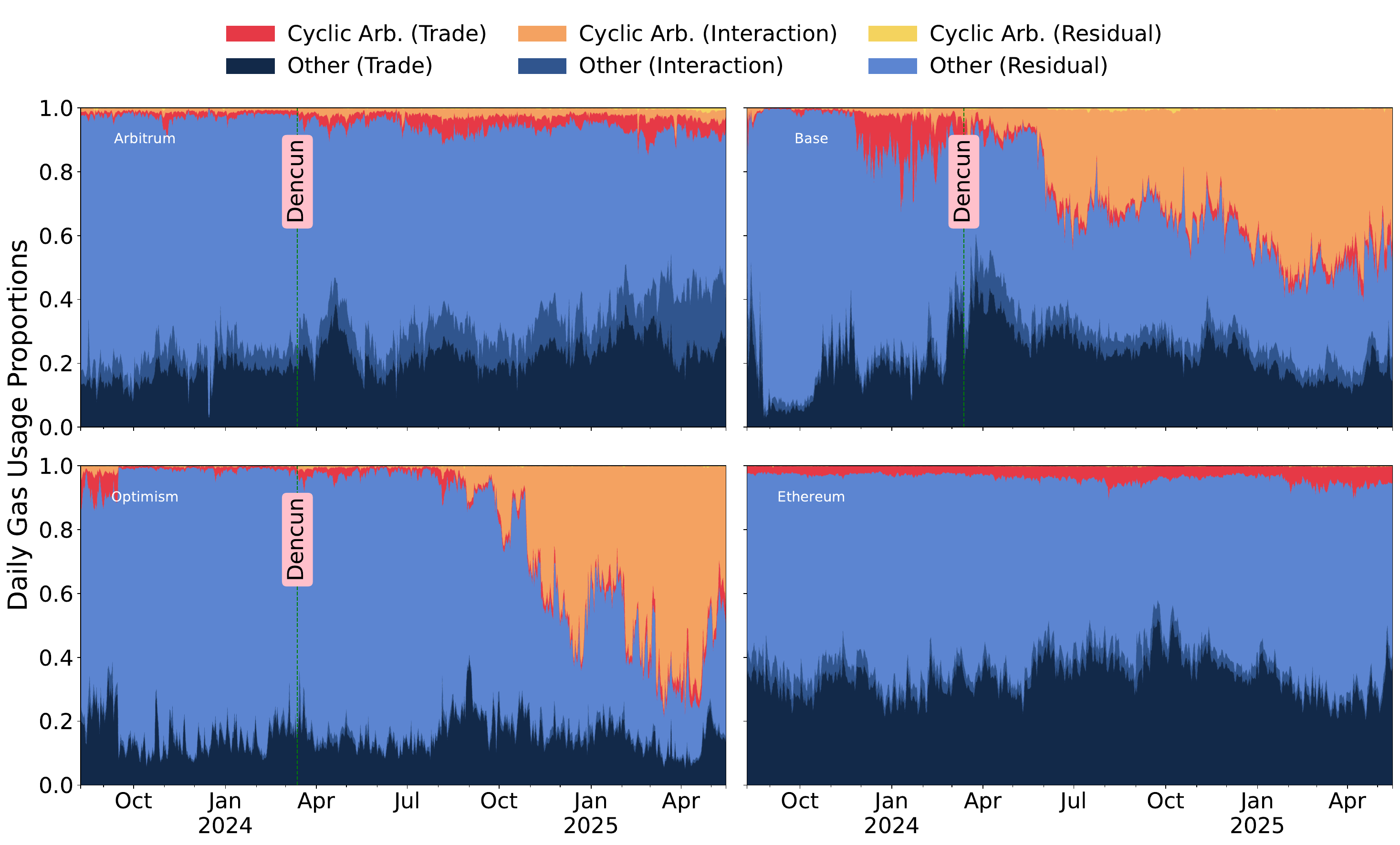}
    \caption{Relative daily gas usage by transaction category on Arbitrum, Base, Optimism, and Ethereum Layer 1. Each area shows gas consumed by cyclic arbitrage MEV bots (\textsc{cyclicArb}) and all other activity (\textsc{other}), subdivided into \textsc{trade} (transactions with executed swaps), \textsc{interaction} (on-chain probing without swaps), and \textsc{residual} segments (see Section~\ref{cat1}). The vertical green dotted line marks the Dencun upgrade (EIP-4844).}
    \label{fig:gas-norm}
\end{figure}

Turning to Figure~\ref{fig:gas-norm}, where we show the relative change in daily gas usage on Arbitrum, Base, Optimism, and Ethereum Layer 1, we focus on the relative usage of the respective categories. Observe that on Base and Optimism \textsc{cyclicArb} - \textsc{interaction} activity surges dramatically after Dencun, quickly outpacing executed arbitrages and consuming the majority of newly available blockspace in 2025. In contrast to the two other Layer 2s, Arbitrum shows no comparable increase in \textsc{cyclicArb} - \textsc{interaction} gas usage, suggesting differing strategic dynamics or protocol-level constraints (e.g., transaction ordering, fees, or shorter interblock times) that may discourage such speculative behavior. Note that Arbitrum displays the largest proportion of \textsc{other} - \textsc{interaction} gas consumption (see Appendix~\ref{app:arbitrum}). This gas usage is driven by multiple contracts, the largest of which is the Uniswap V3 Liquidity Position Manager, but we observe no evidence of cyclic arbitrage from contracts in this category.
Finally, the trade share from probing on the Ethereum Layer 1 remains flat, with trades by non-MEV actors dominant.

Importantly, Figures~\ref{fig:gas} and~\ref{fig:gas-norm} display the categorization without taking the last factor (see \textsc{success} / \textsc{revert} in Section~\ref{cat1}) in consideration. Since optimistic MEV searchers often fail to capture arbitrage opportunities, one might expect a high revert rate among their transactions. However, as shown in Appendix~\ref{revertsapp}, the gas consumed by reverted optimistic MEV transactions is proportionally smaller on Base and Optimism than their share of total gas usage. Thus, on Base and Optimism failed MEV attempts by \textsc{cyclicArb} bots primarily succeed but do not perform trades (i.e., \textsc{cyclicArb} - \textsc{interaction} - \textsc{success}). By contrast, on Arbitrum and Ethereum Layer 1, a substantial share of MEV-related gas goes to reverted transactions, reflecting a higher rate of outright failures (i.e., \textsc{cyclicArb} - \textsc{interaction} - \textsc{revert}). %

Next, we discuss possible factors driving the distinct behavior of optimistic MEV searchers on OP-Stack rollups versus Arbitrum and Ethereum Layer 1. 
First, Arbitrum's FCFS transaction ordering, especially before the Timeboost mechanism,\footnote{TimeBoost, introduced on Arbitrum on April 17, 2025, adds a sealed‐bid second‐price auction ``express lane'' that allows users to submit transactions directly to the sequencer for prioritized inclusion~\cite{arbitrum2025timeboost}.} turns MEV extraction into a latency race, disincentivizing on-chain speculative probing. For example, with FCFS ordering, an optimistic MEV bot cannot specify a low-priority fee to sit at the end of the block and capture price differences after other transactions execute. Instead, it must carefully time its transaction submission, resulting in less control over its block position and over the number of possible preceding transactions that create exploitable price differences.

Second, CEX-DEX arbitrage profits are known to scale with the square root of the mean interblock time~\cite{milionis2023automatedmarketmakingarbitrage, nezlobin2025lossversusrebalancingdeterministicgeneralizedblocktimes}. Given that Arbitrum has shorter block times (approximately 250ms) compared to Base and Optimism (both 2s), the expected profits from CEX-DEX arbitrage are lower on Arbitrum. 
Conversely, the higher expected profits on Base and Optimism from CEX-DEX arbitrage may also elevate profitability in DEX-DEX arbitrage, particularly for pairs not actively traded on centralized exchanges. Once CEX-DEX arbitrage opportunities are closed, residual price discrepancies can remain between DEX pairs, creating additional cyclic arbitrage opportunities within the same Layer 2. Thus, lower block times might discourage excessive transaction spam. 

Finally, block capacity and transaction fees also appear to shape the prevalence of optimistic MEV. In Figure~\ref{fig:gas}, we see that immediately following the Dencun hardfork, Base increased its gas target by roughly an order of magnitude (black dashed line), and Optimism similarly raised its gas target. In contrast, Arbitrum’s speed limit (the equivalent throughput cap) remained unchanged throughout our measurement window.

We conclude by highlighting that in contrast to all three Layer 2s, Ethereum Layer 1 shows almost no significant gas usage in the \textsc{cyclicArb} - \textsc{interaction} category. Instead, gas usage related to DEXes remains dominated by \textsc{other} - \textsc{trade} activity. Activities resembling speculative probing, so prominent on Layer 2s, are largely negligible on Layer 1, reinforcing the fundamental distinction in MEV dynamics between the two environments. Overall, these findings support our core premise: optimistic MEV strategies are uniquely fostered by the low-cost, high-throughput conditions of Layer 2 architectures.

\subsection{Impact of Dencun on Growth}

\begin{figure}[htb]
    \centering
    \includegraphics[width=\linewidth]{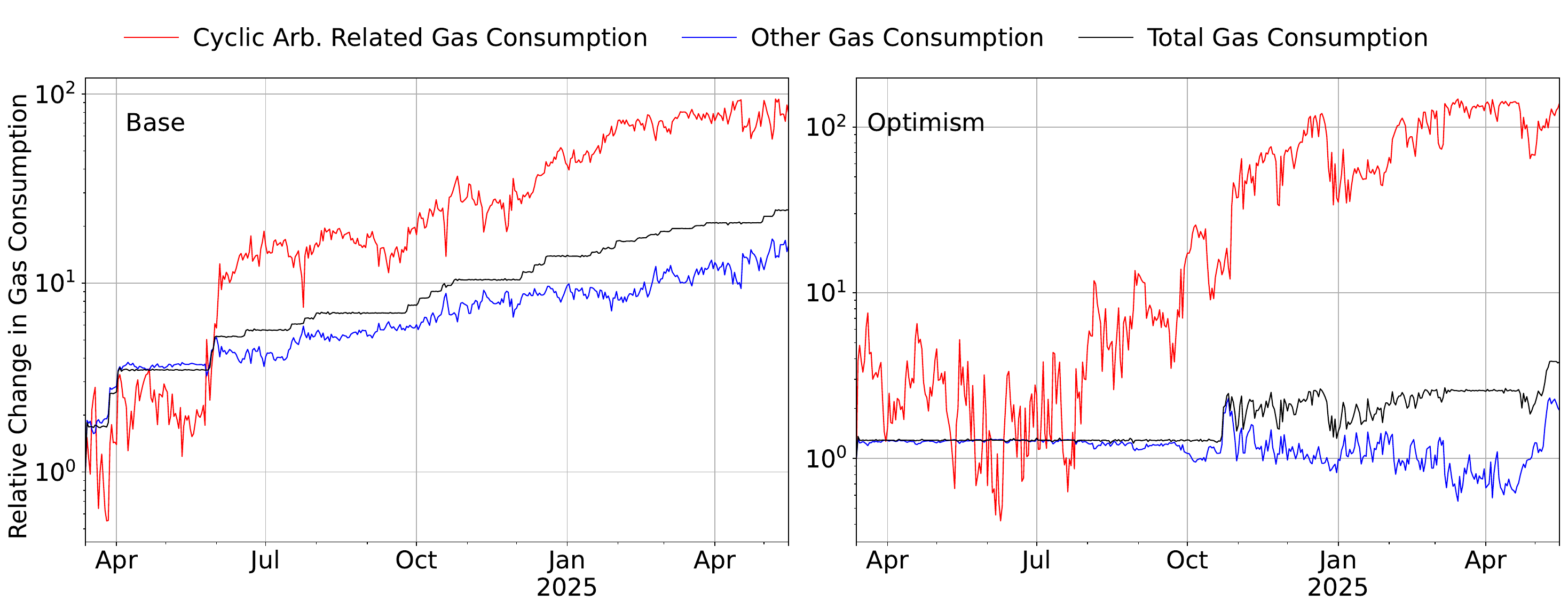}
    \caption{Relative change in daily gas consumption for three categories --- cyclic arbitrage related, other, and total --- on Base and Optimism, normalized to 1 on the day of Dencun hardfork (i.e., 13 March 2024). Each curve shows how gas usage in each category evolves post-Dencun, highlighting the divergent growth rates of optimistic MEV activity (\textsc{cyclicArb}) versus all other on-chain activity.}
    \label{fig:shift}
\end{figure}

To examine this further, Figure~\ref{fig:shift} shows the relative change in daily gas consumption for three categories (i.e., \textsc{cyclicArb}, \textsc{other}, and total activity (\textsc{cyclicArb}$+$\textsc{other})) on Base (left) and Optimism (right), each normalized to 1 on the day of the Dencun upgrade. In both networks, cyclic arbitrage MEV bots exhibit the largest increase, indicating that the additional block capacity is largely consumed by optimistic MEV activity. Notably, on Optimism, there is no consistent rise in gas usage by other transactions despite the higher gas limit. This suggests that the main beneficiary of the additional blockspace is optimistic MEV, with most of the new capacity consumed by on-chain computations that could be done off-chain.

\subsection{Fee Dynamics}
\begin{figure}[htb]
    \centering
    \includegraphics[width=\linewidth]{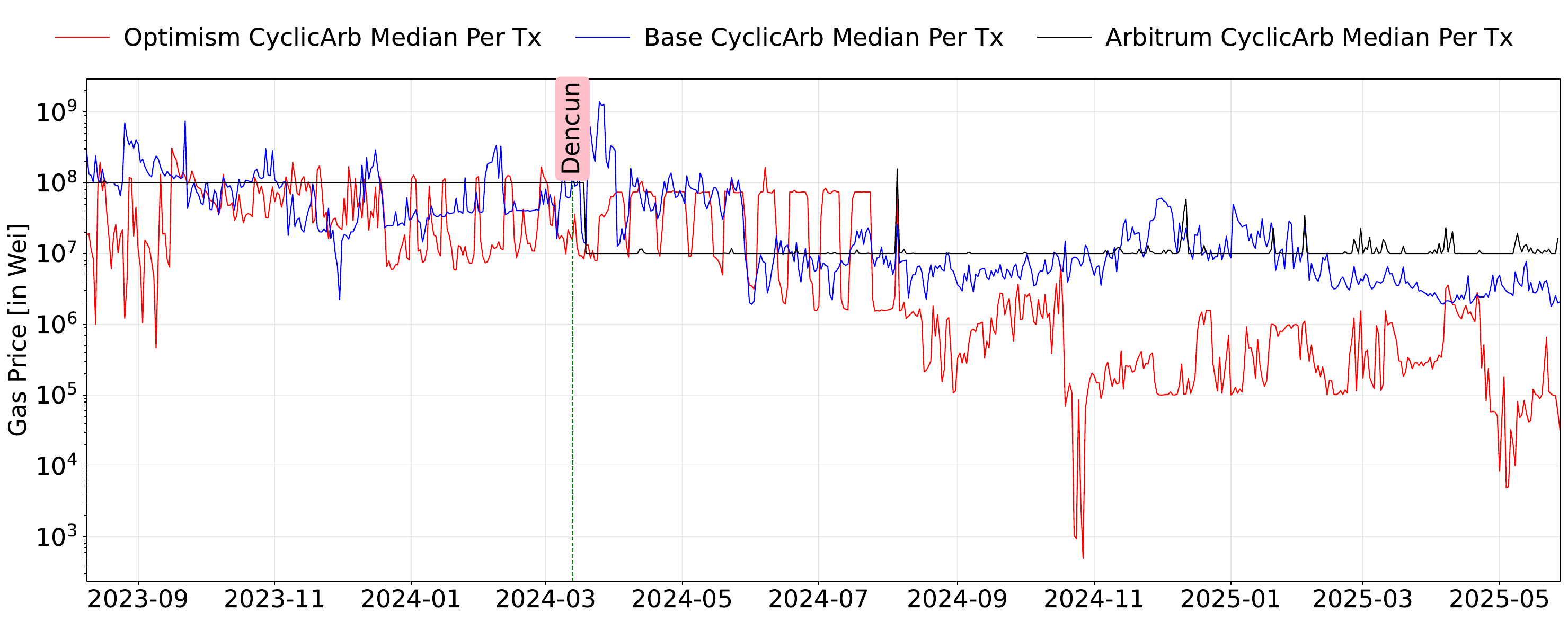}
    \caption{Median gas price paid by optimistic MEV bots (i.e., \textsc{cyclicArb} MEV) on Optimism, Base, and Arbitrum. For Base and Optimism, the plot show \texttt{gas\_price} which includes \texttt{base\_fee} $+$ \texttt{priority\_fee} and for Arbitrum, it shows \texttt{effective\_gas\_price}.}
    \label{fig:gas_price}
\end{figure}

Turning to transaction fees, Figure~\ref{fig:gas_price} plots the daily median gas prices paid by optimistic MEV bots on each Layer 2. Notably, Arbitrum enforces a floor of 0.01~GWei on its base fee when demand falls below its speed limit \cite{arbitrumgasfees}. In contrast, Base and Optimism follow the EIP-1559 model, allowing fees to decline arbitrarily when demand is under the gas target. This difference explains why Base and Optimism blocks appear consistently full (see Figure~\ref{fig:gas}) in comparison to target utilization, whereas Arbitrum’s blocks often show unused headroom. 

Median gas prices paid by optimistic MEV bot transactions (i.e., those labeled as \textsc{cyclicArb} MEV) starting from Dencun, further underscore these distinctions: Optimism sees the lowest median price (0.0005~GWei), Base sits in the middle (0.0061~GWei), and Arbitrum commands the highest median (0.01~GWei). Thus, the median price paid by optimistic MEV bots on Optimism is a factor of 20 lower than on Arbitrum, while it is nearly a factor of 2 lower on Base than on Arbitrum.\footnote{One reason the gap between Base and Arbitrum is smaller is that there is a short period where fees on Base are higher than those on Arbitrum. Between late November 2024 and early January 2025, we see the gas price paid by optimistic MEV bots on Base exceeding Arbitrum. The reason for this is the congestion in the network during those periods caused by the sniper bots (see Appendix~\ref{revertsapp}).} Lower fees on Base and Optimism reduce the cost of speculative probing, making optimistic MEV more profitable, while Arbitrum’s relatively higher fee floor dampens such behavior. Thus, a final factor driving optimistic MEV is the comparatively low fees on Base and Optimism. These reduced costs make speculative probing more profitable on those networks, whereas Arbitrum’s higher fee floor suppresses such behavior.

Importantly, even though transactions from optimistic MEV searchers on Base and Optimism account for more than half the gas usage in the first quarter of 2025, they only account for 23\% and 17\% of the transaction fees paid, respectively. Thus, there is a disconnect between the gas used and the fees paid. Optimistic MEV transactions pay less for their transactions. 

\begin{figure}[htb]\vspace{-10pt}
    \centering
    \includegraphics[width=\linewidth]{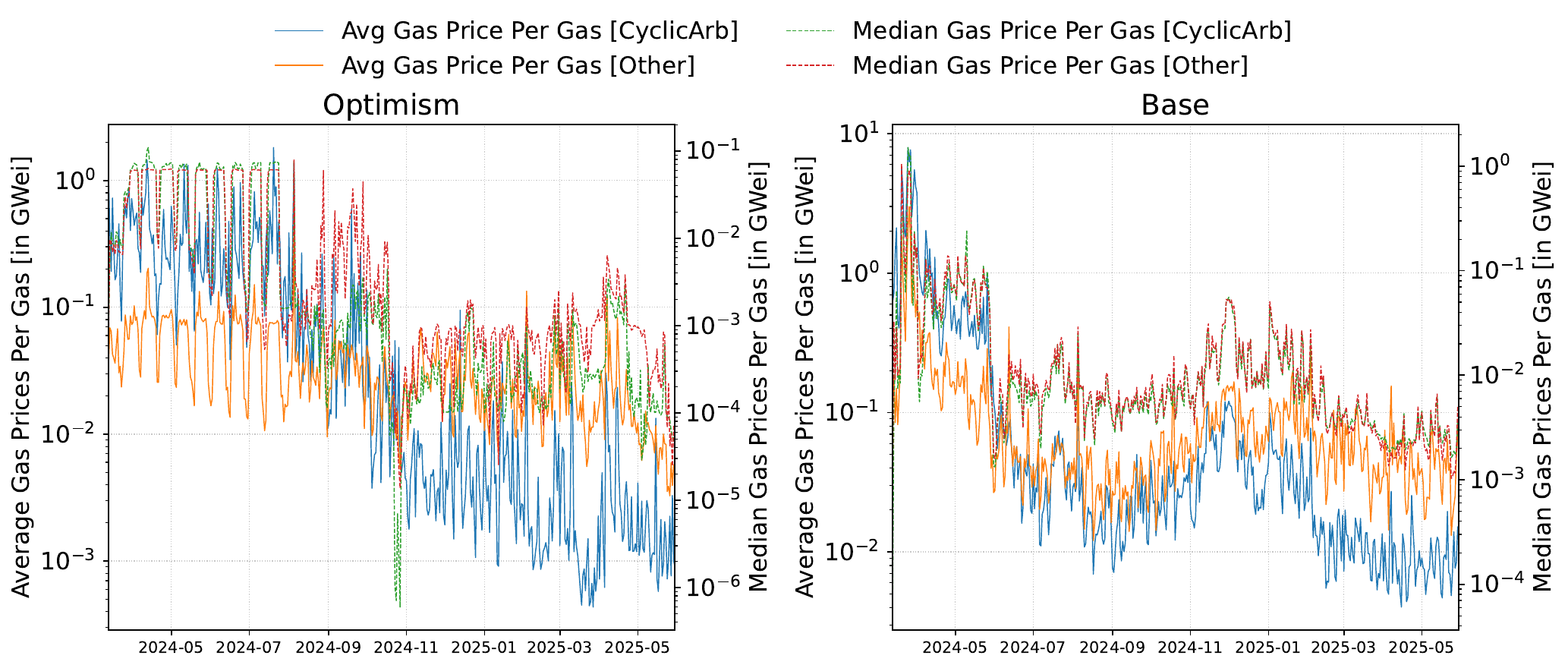}
    \caption{Gas prices per gas unit from optimistic MEV bots (i.e., \textsc{cyclicArb} MEV) versus other actors on Base (right) and Optimism (left) starting from the day of Dencun hardfork (i.e., 13 March 2024). Solid lines show the daily average gas price per gas unit paid by \textsc{cyclicArb} MEV bots (blue) and other actors (orange), while dashed lines show the corresponding median fees. Both average and median gas prices paid per gas unit by \textsc{cyclicArb} MEV bots are lower than those of others.}
    \label{fig:gas1}
\end{figure}

Next, we examine this fee discrepancy in more detail in Figure~\ref{fig:gas1}, which plots median and mean gas price per unit gas paid by optimistic MEV bots (i.e., those classified as \textsc{cyclicArb}) versus all other transactions on Base and Optimism over time.

On Base, in Q1 2025, the average gas price per gas paid by optimistic MEV bots is 0.0209~GWei, compared with 0.0710~GWei for all other transactions --- a factor of 3.5 difference. When we consider the median instead of the mean, the gap narrows: MEV bots pay a median of 0.0047~GWei, while the rest pay 0.0057~GWei. Notably, between April and September 2024, there was a period when MEV bot transactions actually paid higher average fees than other users. This shows that, although optimistic MEV bots tend to pay less overall, the difference is smaller and more nuanced than one might expect.

On Optimism, the pattern is similar. In Q1 2025, the average gas price per gas for MEV bots is 0.0042~GWei, versus 0.0252~GWei for other actors --- a factor of 5 difference. Median gas prices are 0.0003~GWei for MEV bots and 0.0007~GWei for the rest. Again, the smaller median gap and occasional fee spikes by MEV bots in mid-2024 confirm that on-chain probing does not always seek the absolute cheapest gas. Nonetheless, the high prevalence of optimistic MEV correlates with periods of low gas prices after the Dencun hardfork, as we saw previously.

\subsection{Characteristics of the Largest Cyclic Arbitrage Bots}

We now turn our attention to the behavior and structure of the most active \textsc{cyclicArb} bots on each network. Specifically, we analyze the top 10 bots by gas usage on Base, Optimism, and Arbitrum, focusing on execution outcomes, code similarity, and operational architecture.

The statistics relevant to the top 10 bots (based on gas usage) are presented in the following tables for Base (Table~\ref{tab:mev_bot_activity_base}), Optimism (Table~\ref{tab:mev_bot_activity_optimism}), and Arbitrum (Table~\ref{tab:mev_bot_activity_arbitrum}). The primary metrics in the tables are defined as follows:
\begin{description}
    \item[Swaps] The total number of individual swap operations executed as a result of calls to the given MEV bot contract. Notably, a single on-chain transaction may include multiple distinct swap operations.
    \item[Transactions With Trades] The total number of unique transactions that called the MEV bot contract and included one or more swap operations. It counts transactions rather than individual swaps.
    \item[Non-Reverted Transactions] The total number of transactions that called the MEV bot contract and were successfully executed and committed on-chain.
    \item[Reverted Transactions] The total number of transactions that called the MEV bot contract but failed during execution (i.e., were reverted), and therefore did not result in the intended state changes, although they still consumed gas.
    \item[Transaction per Unique Calldata] The ratio total number of transactions by the bot to the unique calldata entries used, where a value of 10 means that, on average, each calldata entry is reused across 10 transactions.
    \item[Median Calldata Length] The median length of the calldata used by the bot (in hexadecimal characters).
    \item[Cumulative MEV Bot Gas (\%)] The cumulative percentage of total transaction gas consumed by MEV bot contracts. The percentage is calculated relative to the total gas used by a defined set of contracts (i.e., all identified MEV bots within the dataset $\mathcal{C}'_{bots,chain}$) and is rounded to two decimal places.
    
\end{description}

\begin{table}
\begin{subtable}[h]{\textwidth}
  \centering
  \setlength{\tabcolsep}{3pt} %

  \centering

  \setlength{\tabcolsep}{3pt} %
  \sisetup{
  }
  \resizebox{1\textwidth}{!}{

  \begin{tabular*}{1.2\linewidth}{@{} 
    l %
    S[table-format=9, group-separator={,}]  %
    S[table-format=9, group-separator={,}]  %
    S[table-format=10, group-separator={,}]  %
    S[table-format=8, group-separator={,}]       
    S[table-format=10, group-separator={,}]      
    S[table-format=8, group-separator={,}]                     
    S[table-format=10, round-mode=places, round-precision=2, table-space-text-post={\,\%}]
    @{}
  }
    \toprule
    {contract} & %
    {swaps} &    %
    {\shortstack{txs with\\trades}} &   %
    {\shortstack{non-reverted \\txs}} &  %
    {\shortstack{reverted \\txs}} &  %
    {\shortstack{txs per\\ unique calldata}}&
    {\shortstack{median\\ calldata length}} &
    {\shortstack{cum. MEV \\ bot gas (\si{\%})}}\\ %
    \midrule
    \href{https://basescan.org/address/0xf5ff765b0c1278e54281193d7019281e0e50a8c0}{\texttt{0xf5ff\dots a8c0}} & 175301 &  60315 & 23173896 &    1  & 107.08 & 3830&  6.71 \\
    \href{https://basescan.org/address/0xbff69c6f5bb902807b05d0d1a523e3b20b3b73e9}{\texttt{0xbff6\dots 73e9}} & 411376 & 152046 & 24379515 & 3700  & 90.95 & 2977& 11.76\\
    \href{https://basescan.org/address/0xaa877b9d9fbd4d6c40c3b9503f0d251531f1fb84}{\texttt{0xaa87\dots fb84}} & 128100 &  65074 & 14680490 &  535  & 16.65 & 2722& 15.16\\
    \href{https://basescan.org/address/0xddfa17794fdd2903f16a496ebefb76a87355abd3}{\texttt{0xddfa\dots abd3}}  &  42715 &  15948 &  9032099 &   84  & 3.14 & 1255& 17.90\\
    \href{https://basescan.org/address/0xdade45a1acda95c981040aaedb6fe30cf3bc6084}{\texttt{0xdade\dots 6084}} & 148186 &  51864 &  5338322 &   28  & 145.06 & 6 & 19.42\\
    \href{https://basescan.org/address/0xe91ca2ee0566e76c99321435e9df5817b9c996af}{\texttt{0xe91c\dots 96af}} &  56531 &  18637 &  6880362 &  379 & 70.72 & 1330 & 20.50 \\
    \href{https://basescan.org/address/0x4e85efbb92d53d1ce0e8a0e5129b881c5ffa0cea}{\texttt{0x4e85\dots 0cea}} & 125575 &  45586 &  1765057 &   25  & 44.43 & 2& 21.56\\
    \href{https://basescan.org/address/0x2b24051780aeea28aec159ae8e46528fafe5446f}{\texttt{0x2b24\dots 446f}}  &  65548 &  22627 &  3542945 &    9 & 129.45 & 6  & 22.58\\
    \href{https://basescan.org/address/0x826fd727477547bd89d75f7941d35f525c04b5f5}{\texttt{0x826f\dots b5f5}} & 343999 & 106016 &  5009861 &    3 & 1.51 & 2465& 23.58  \\
    \href{https://basescan.org/address/0xbba96510886d6abbb917da35add5362cd8e33cf9}{\texttt{0xbba9\dots 3cf9}} &  68352 &  25976 &  2737139 &   14  & 76.01 & 2& 24.56 \\
    \bottomrule
  \end{tabular*}}\vspace{3pt}
  
  \caption{Base}
  \label{tab:mev_bot_activity_base}
\end{subtable}

\begin{subtable}[h]{\textwidth}
  \centering

  \setlength{\tabcolsep}{3pt} %
  \sisetup{
  }
  \resizebox{1\textwidth}{!}{
  \begin{tabular*}{1.2\linewidth}{@{} 
    l %
    S[table-format=9, group-separator={,}]  %
    S[table-format=9, group-separator={,}]  %
    S[table-format=10, group-separator={,}]  %
    S[table-format=8, group-separator={,}]     
    S[table-format=10, group-separator={,}] 
    S[table-format=8, group-separator={,}]                     
    S[table-format=10, round-mode=places, round-precision=2, table-space-text-post={\,\%}]
    @{}
  }
    \toprule
    {contract} & %
    {swaps} &    %
    {\shortstack{txs with\\trades}} &   %
    {\shortstack{non-reverted \\txs}} &  %
    {\shortstack{reverted \\txs}} &  %
    {\shortstack{txs per\\ unique calldata}}&
    {\shortstack{median\\ calldata length}} &
    {\shortstack{cum. MEV \\ bot gas (\si{\%})}}\\ %
    \midrule
\href{https://optimistic.etherscan.io/address/0x887290c34856cd3ba1b84da78cccf43812f66324}{\texttt{0x8872\dots 6324}} & 676753& 156634 & 11966203 & 0  &498.11 & 6& 33.06\\
\href{https://optimistic.etherscan.io/address/0xabf4daac18925530d1e4f99fd538d57b8bf1017c}{\texttt{0xabf4\dots 017c}}& 29586 & 8251 & 7979548 & 0 & 7979548.00 &10 & 41.94\\
\href{https://optimistic.etherscan.io/address/0xcdcca22e3b1e0170532eea4b14e5fb3a73b45c8f}{\texttt{0xcdcc\dots 5c8f}} & 247308 & 68040 & 6382847 & 0  & 3489.80 & 6& 49.08\\
\href{https://optimistic.etherscan.io/address/0xd3dc079ac6f98275bbbcd0aff11cbaadb4d8f2ac}{\texttt{0xd3dc\dots f2ac}} & 214741 & 60522 & 6901973 & 5  & 1674.42 &  6 & 55.96\\
\href{https://optimistic.etherscan.io/address/0x4d43aa8a0d161a7f2e7ba6c4226fe67998ee987f}{\texttt{0x4d43\dots 987f}} & 208002 & 47459 & 2322899 & 10 & 195.39 & 3213 & 62.34\\ 
\href{https://optimistic.etherscan.io/address/0x3955945c12eb164d2356de1a290ec54fa4c574c2}{\texttt{0x3955\dots 74c2}} & 422807 & 107201 & 1992783 & 0  & 2.98 & 2& 64.56 \\ 
\href{https://optimistic.etherscan.io/address/0x9d1b033ac8bff2b07fb7d13385b8c270db25f96f}{\texttt{0x9d1b\dots f96f}} & 37920 & 9531 & 825165 & 40  &  1650.41 & 6& 66.62\\ 
\href{https://optimistic.etherscan.io/address/0x0daf895a78eb49151a5f1003818939770d3ca7dd}{\texttt{0x0daf\dots a7dd}} & 398174 & 105069 & 2299930 & 0  & 4.72 & 2 & 68.51\\
\href{https://optimistic.etherscan.io/address/0x2642f2b26cce1b96f2a94ee1660d5e7c7b9839e6}{\texttt{0x2642\dots 39e6}} & 73322 & 26656 & 1387723 & 242  & 775.83 & 2422& 69.98\\
\href{https://optimistic.etherscan.io/address/0xf2610cca617c94d3b617ec5872d191156bcfc883}{\texttt{0xf261\dots c883}} & 242058 & 51232 & 928822 & 4004  & 100.14 & 6& 71.44\\
    \bottomrule
  \end{tabular*}}\vspace{3pt} 
  
  \caption{Optimism}
  \label{tab:mev_bot_activity_optimism}
\end{subtable}

\begin{subtable}[h]{\textwidth}
  \centering
  \setlength{\tabcolsep}{5pt} %
  \sisetup{
  }
  \resizebox{1\textwidth}{!}{
  \begin{tabular*}{1.27\linewidth}{@{} 
    l %
    S[table-format=9, group-separator={,}]  %
    S[table-format=9, group-separator={,}]  %
    S[table-format=10, group-separator={,}]  %
    S[table-format=8, group-separator={,}]                     
    S[table-format=10, group-separator={,}] 
    S[table-format=8, group-separator={,}]                     
    S[table-format=10, round-mode=places, round-precision=2, table-space-text-post={\,\%}]
    @{}
  }
    \toprule
    {contract} & %
    {swaps} &    %
    {\shortstack{txs with\\trades}} &   %
    {\shortstack{non-reverted \\txs}} &  %
    {\shortstack{reverted \\txs}} &  %
    {\shortstack{txs per\\ unique calldata}}&
    {\shortstack{median\\ calldata length}} &
    {\shortstack{cum. MEV \\ bot gas (\si{\%})}} \\ %
    \midrule
\href{https://arbiscan.io/address/0x00000000cfe3369bcdbc76071ba6e0a4e0fe98bd}{\texttt{0x0000\dots 98bd}} & 9871348 & 3338292 & 3338353 & 1241384  & 17.34& 110& 10.42\\
\href{https://arbiscan.io/address/0x60ca5def792e7528f8ff48b7aea60c1e2234294b}{\texttt{0x60ca\dots 294b}} & 5106002 & 1836869 & 4355290 & 26968  & 1.12 & 1207& 18.68\\
\href{https://arbiscan.io/address/0x689368aae04fe7d961d1e545b28266c4f179151c}{\texttt{0x6893\dots 151c}} & 4769221 & 1706662 & 1708167& 1087474  & 1.00 & 572& 24.28\\
\href{https://arbiscan.io/address/0x9e52272477c2fcbe2c38bf12dd06cb5739975867}{\texttt{0x9e52\dots 5867}} & 3055148 & 1065722 & 1067478 & 1317339  & 1.00 & 336 & 29.22\\
\href{https://arbiscan.io/address/0xa9ff271ee217dc1c9ce3f7ebf0d6f096842cd82f}{\texttt{0xa9ff\dots d82f}} & 774311 & 347885 & 348021 & 51732  & 1.05 & 2353& 32.67\\
\href{https://arbiscan.io/address/0x00000000c868e634590a981dd3bd007dca5c2e43}{\texttt{0x0000\dots 2e43}} & 3045007 & 1125161 & 1125162 & 338202  & 14.52 & 104& 35.34 \\
\href{https://arbiscan.io/address/0x0000b0ca00000017d543c800be4257629544fb51}{\texttt{0x0000\dots fb51}} & 1527676 & 502973 & 503320 & 290686  & 1.95 & 163& 37.53\\
\href{https://arbiscan.io/address/0xf238d4353246948ecdc3f3252ae939fe5234e145}{\texttt{0xf238\dots e145}} &  320724 & 116712 & 585063 & 2796  & 2.39 & 266& 39.28\\ 
\href{https://arbiscan.io/address/0xe98b3b8d43842a52efc848bf10be28d65c2ed87c}{\texttt{0xe98b\dots d87c}} & 1145798 & 560741 & 516379 & 413451& 30.23 & 52 & 40.87 \\
\href{https://arbiscan.io/address/0x84f1d597f1160918d9ec28c5ab0b2eb5394bb7cf}{\texttt{0x84f1\dots b7cf}} & 522016 & 248810 & 252743 & 251681 & 17.31 & 46& 42.46 \\
    \bottomrule
  \end{tabular*}}\vspace{3pt}
  
  \caption{Arbitrum}
  \label{tab:mev_bot_activity_arbitrum} 
\end{subtable}
\caption{This table reports statistics for the top 10 cyclic arbitrage MEV bot contracts (ranked by cumulative gas usage) on Base (Table~\ref{tab:mev_bot_activity_base}), Optimism (Table~\ref{tab:mev_bot_activity_optimism}), and Arbitrum (Table~\ref{tab:mev_bot_activity_arbitrum}). For each contract, we report: the total number of individual swap operations executed (swaps), the number of transactions that included at least one swap (txs with trades), the number of transactions that executed without reverting (non-reverted txs), and the number of transactions that reverted during execution (reverted txs). The column txs per unique calldata gives the ratio of total transactions to the number of unique calldata entries used by the bot and median calldata length reports the median size of calldata (in hexadecimal characters) observed for the bot. The last column (cum. MEV bot gas (\si{\%})) indicates the cumulative share of gas consumed by each contract relative to all identified MEV bots on that chain.} \label{tab:mev_bot_activity}
\end{table}
\noindent

\paragraph*{Execution Outcomes and Bot Behavior Across Networks}

A principal finding of our analysis concerns the execution outcomes of cyclic arbitrage attempts by MEV bots. Across all three Layer 2 networks, i.e., Base, Optimism, and Arbitrum, we observe that a large proportion of cyclic arbitrage attempts do not result in profitable outcomes as they do not even execute a swap. However, the way unsuccessful attempts are handled differs substantially between networks.

On OP-Stack-based networks (Base and Optimism), many unprofitable MEV transactions still conclude successfully from the perspective of the \textit{Ethereum Virtual Machine (EVM)}; that is, they do not revert (on Optimism revert rate is 0.01\%, on Base 0.005\%), even though they fail to execute a profitable arbitrage. These transactions typically make no substantive state changes apart from incurring transaction fees. This behavior implies that the primary disincentive for speculative probing on these chains is economic (i.e., fees), not technical failure. Despite their technical success, our analysis shows that the actual success rate (defined as executing an arbitrage) is exceptionally low among the top contracts on Base (0.58\%) and Optimism (1.49\%). This indicates a high volume of ultimately unsuccessful, speculative execution.

In contrast, Arbitrum displays a different pattern. When transactions initiated by the top cyclic arbitrage bots do not revert, they are overwhelmingly likely to result in a successful arbitrage. There, the success rate of non-reverted transactions is 77\%, while the overall success rate (i.e., including those transactions that revert) is 56\%. This suggests a more selective or precise execution strategy. However, this high success rate for non-reverted transactions is accompanied by a significantly higher incidence of reverts, relative to Base and Optimism. Thus, Arbitrum-based bots appear to favor failing fast (via reverts) over speculative probing. These findings were further validated through cross-referencing with a curated dataset prepared by Entropy Advisors, an official partner of the Arbitrum Foundation~\cite{EntropyAdvisors2025,EntropyDune2025}, strengthening our confidence in the observed execution dynamics across networks.

Further, our analysis of the largest cyclic arbitrage bots demonstrates again that they predominantly operate as optimistic MEV bots. To demonstrate this, we compute the number of transactions per unique calldata entry. On Base, the largest cyclic arbitrage bot reuses the same calldata more than 100 times on average, while on Optimism, one bot repeats identical calldata over 7~million times. Across both Base and Optimism, all major cyclic arbitrage contracts exhibit extensive calldata reuse, reflecting a speculative approach where little specific information such as exact paths or trade amounts is precomputed off-chain. This is indicative of optimistic MEV. Examining the calldata itself reinforces this: in many cases it is non-informative, as indicated by short median calldata lengths, or consists primarily of lists of DEXes to probe for potential opportunities.

When we turn to Arbitrum, the picture is somewhat different. While the majority of the largest cyclic arbitrage contracts exhibit extensive calldata reuse, several do not. In particular, the third, fourth, and fifth largest contracts show minimal reuse, resembling the pattern observed on Ethereum, where none of the major cyclic arbitrage bots have a transactions per unique calldata ratio higher than 1.006. Despite this, their transactions revert frequently, around 50\%, suggesting they are not speculative in the sense of performing computation on-chain but are instead exposed to significant uncertainty about whether their transactions will land as intended. The remaining large cyclic arbitrage bots on Arbitrum, however, display frequent calldata reuse, consistent with the optimistic MEV pattern observed on Base and Optimism.

\paragraph*{Concentration of Gas Usage}

We also compared the concentration of MEV bot activity across networks in terms of gas usage. On Optimism, the top 10 MEV bots account for a disproportionately large share of gas consumed by atomic arbitrage, suggesting a more concentrated and potentially less competitive landscape. Arbitrum shows a slightly more distributed gas usage among top bots, while Base has the most diffuse distribution. These trends align with recent shifts in dominance across Layer 2s, with Base emerging as the most active MEV network, followed by Arbitrum and then Optimism.

\paragraph*{Code Similarity and Cloning Behavior}
To explore potential relationships among MEV-performing contracts, we conducted a bytecode similarity analysis inspired by prior work on Ethereum contract topology by Kiffer et al.~\cite{10.1145/3278532.3278575}. This revealed meaningful differences in code reuse and cloning behavior across networks.

Contracts deployed on Optimism and Base show a moderate degree of bytecode similarity, suggesting some shared tooling or deployment patterns. In contrast, cyclic arbitrage contracts on Arbitrum appear more structurally distinct, with fewer instances of directly shared or closely related implementations. Among the three networks, Optimism exhibits the highest level of intra-network code similarity, pointing to a relatively concentrated MEV ecosystem with repeated use of identical or near-identical contracts --- potentially indicating lower competitive diversity.

We also identified clear examples of contract cloning behavior, particularly on Base (see Appendix~\ref{app:clone}). For instance, contracts deployed at addresses \href{https://basescan.org/address/0x2b24051780aeea28aec159ae8e46528fafe5446f#code}{\texttt{0x2b24...446f}} and \href{https://basescan.org/address/0xdade45a1acda95c981040aaedb6fe30cf3bc6084}{\texttt{0xdade...6084}} share identical bytecode not only with each other but also with 48 additional contracts. All 50 of these contracts engage in cyclic arbitrage and are classified as MEV bots in our dataset~\cite{BaseScan2025sim50}. Although none of these clones individually ranked among the top 10 gas consumers, their collective presence underscores the widespread use of standardized MEV bot implementations. This pattern may reflect coordinated deployments by a single actor or group.

Further analysis of these cloned contracts reveals another striking pattern: the transaction \texttt{calldata} passed to them is often minimal or non-informative (e.g., \texttt{0x0001}), suggesting that key arbitrage parameters are not set off-chain. As an example, \href{https://basescan.org/address/0xdade45a1acda95c981040aaedb6fe30cf3bc6084}{\texttt{0xdade...6084}} executed 51,864 transactions with trades of which 50,861 had \texttt{0x0001} as calldata. Instead, the contract logic itself likely handles opportunity identification, path selection, and trade sizing through real-time on-chain computation. This architecture relies heavily on internal heuristics and on-chain state queries (e.g., via \texttt{STATICCALL}s), contrasting with more traditional MEV bots where critical parameters are calculated off-chain and passed in via calldata.

\section{Drivers of Optimistic MEV}

To move beyond descriptive patterns and develop a quantitative understanding of the drivers of MEV activity on Layer 2 networks, we employ an OLS regression framework. The goal is to assess the statistical significance and direction of influence of various market and network-level metrics on daily MEV fluctuations. By systematically analyzing these relationships, we aim to identify the conditions under which MEV activity is amplified or suppressed across evolving Layer 2 ecosystems.

Our analysis focuses on two dependent variables: the daily change in the total number of transactions associated with \textsc{cyclicArb} bots ($\Delta$\texttt{CyclicArbTx}$_t$) and the subset of those transactions that execute trades ($\Delta$\texttt{CyclicArbTxWTrade}$_t$), described in detail below:

\begin{description}
    \item[$\Delta \text{CyclicArbTx}_t$ (Change in Cyclic Arbitrage Transaction Count)]
    The total number of transactions classified as \textsc{cyclicArb} (see Section~\ref{cat1}) within our dataset is measured by this variable. 
    \item[$\Delta \text{CyclicArbTxWTrade}_t$ (Change in Cyclic Arbitrage Trade Transaction Count)]
    The daily change in the total number of transactions that are classified as \textsc{cyclicArb}-\textsc{trade}. 
\end{description}

This dual perspective allows us to distinguish between general \textsc{cyclicArb} activity and the subset of transactions involving explicit on-chain value extraction through token trades, \textsc{cyclicArb}-\textsc{trade}. Understanding these dynamics is crucial for assessing the efficiency of Layer 2 markets and informing the design of potential mitigation strategies.

The regression equation is then specified as:
\begin{align*}
\Delta y_t ={} \beta_0 &+ \beta_1 \cdot \Delta \text{Price}_t + \beta_2 \cdot \Delta \text{Volatility}_t 
               + \beta_3 \cdot \Delta \text{RetailTxs}_t \\
              & + \beta_4 \cdot \Delta \text{RetailAggFrac}_t 
               + \epsilon_t
\end{align*}

where $\Delta x_t = x_t - x_{t-1}$ denotes the first difference of variable $x$ at time $t$, and $\epsilon_t$ is the error term. The independent variables are defined as follows:
\begin{description}
    \item[$\Delta \text{Price}_t$ (Change in ETH Price)] The daily change in the price of ETH (denominated in US\$).
    \item[$\Delta \text{Volatility}_t$ (Change in ETH Volatility)] The daily change in intraday volatility of ETH. Intraday volatility is computed using the Garman-Klass estimator~\cite{gk1,li2021intraday}, defined for a given day as:
    $$\sigma_{\text{OHLC}} = \sqrt{0.5 \cdot (\ln H - \ln L)^2 - (2\ln2 - 1) \cdot (\ln C - \ln O)^2}$$
    where $O, H, L, C$ are the Open, High, Low, and Close prices for ETH within the day, respectively.
    \item[$\Delta \text{RetailTxs}_t$ (Change in Retail Trade Count)] The daily change in the total number of on-chain trades --- defined as transactions involving at least one swap --- initiated by entities not classified as \textsc{cyclicArb} bots in our curated dataset, i.e., \textsc{other}-\textsc{trade}.
    \item[$\Delta \text{RetailAggFrac}_t$ (Change in Retail Aggregator Usage)]  
The daily change in the fraction of trades by non-cyclic arbitrage bot addresses that are routed through DEX aggregators:

$$
\text{RetailAggFrac}_t = \frac{\text{RetailAggregatorTrades}_t}{\text{RetailTxs}_t}
$$

A lower value indicates a greater share of direct-to-pool trades, which may reflect routing inefficiencies and create arbitrage opportunities~\cite{angeris2022optimalroutingconstantfunction,thogardpvptweet2022a,thogardpvptweet2022b}. However, not all direct trades are necessarily exploitable, some may avoid introducing new arbitrage opportunities and thus remain economically efficient, even without using an aggregator.

This metric excludes trades routed via standard Uniswap routers, which differ in function and behavior from third-party DEX aggregators~\cite{hackmduniswapbug,mevrefundtweet2023,foldfinancetweet2022,uniswapuniversalroutergithub,uniswapuniversalrouteroverview}.
\end{description}

We now turn to the empirical results of our regression analysis, presented in Table~\ref{tab:ols_results}. The models examine how the previously defined variables relate to daily fluctuations in \textsc{cyclicArb} activity across Layer 2 networks. Specifically, we report results for two dependent variables: the daily change in overall \textsc{cyclicArb}-related transactions ($\Delta \texttt{CyclicArbTx}$) and the daily change in \textsc{cyclicArb}-related transactions involving DEX trades ($\Delta \texttt{CyclicArbTxWTrade}$).

\begin{table}[htb]
\centering
\resizebox{1\textwidth}{!}{
\begin{tabular}{@{}lcccccc@{}}
\toprule
& \shortstack{$\Delta$CyclicArbTx \\ (1, Base)} & \shortstack{$\Delta$CyclicArbTxWTrade \\ (2, Base)} & \shortstack{$\Delta$CyclicArbTx \\ (1, Optimism)} & \shortstack{$\Delta$CyclicArbTxWTrade \\ (2, Optimism)} & \shortstack{$\Delta$CyclicArbTx \\ (1, Arbitrum)} & \shortstack{$\Delta$CyclicArbTxWTrade \\ (2, Arbitrum)}\\
\midrule
const & 2445.7691 & 173.5872 & 372.2486 &  44.9657 & 82.2414 & 32.3800\\
& (6290.7125) & (582.2378) & (1066.6453) & (152.2307)& (437.4133) & (177.3206) \\
$\Delta $Price & -12608.7057 & -1881.3146$^{**}$ & -1156.6053 & -454.8404 & -25.9810 & -357.1149  \\
& (10629.4760) & (897.7109) & (1421.6869) & (298.1963)& (706.7666) & (264.1559) \\
$\Delta$Volatility & 34112.7533$^{***}$ & 8919.1607$^{***}$ & 1323.4766 & 1289.5335$^{***}$ & 2265.7813$^{***}$ & 299.0799\\
& (11474.9164) & (913.0034) & (1722.3345) & (375.0010) & (707.9527) & (307.2508)\\
$\Delta$RetailTxs & 23175.6525$^{**}$ & 4172.5237$^{***}$ & 2956.2109 & 3276.7001$^{***}$ & 16024.0033$^{***}$ & 8097.8011$^{***}$\\
& (10917.7060) & (1245.3633) & (1837.3891) & (532.7932)& (904.0916) & (360.9759) \\
$\Delta$RetailAggFrac & 12230.8890$^{**}$ & 1883.1927$^{***}$ & -1785.4280 & -1264.8155$^{***}$ & -1524.6186$^{***}$ & -1319.3665$^{***}$\\
& (4756.1076) & (442.2920) & (1253.4087) & (369.1399)& (475.3803) & (289.9207)\\
\midrule
Obs & 654 & 654 & 700 & 700 &  700 & 700 \\
Adj. $R^2$ & 0.0736 & 0.3399 &  0.0240 & 0.5818& 0.7143 & 0.7898\\
F-stat & 6.8461 & 31.7334& 2.7295 & 57.8104& 208.3965 & 265.6485  \\
\bottomrule
\end{tabular}
}\vspace{3pt}

\caption{OLS regression results for Base, Optimism, and Arbitrum. The dependent variables are the daily change in \textsc{cyclicArb}-related transactions ($\Delta$CyclicArbTx) and the daily change in \textsc{cyclicArb}-related transactions involving DEX trades ($\Delta$CyclicArbTxWTrade). Independent variables include changes in ETH price, intraday volatility, the count of non-\textsc{cyclicArb} (\textsc{other}, see Section~\ref{cat1}) trades, and the fraction of those routed via DEX aggregators. Robust standard errors are reported in parentheses. Significance levels: $^{*}p<0.1$, $^{**}p<0.05$, $^{***}p<0.01$.}
\label{tab:ols_results}
\end{table}

A consistent pattern across all three networks is that the model explains daily changes in \textbf{\textsc{cyclicArb}-\textsc{trade} transaction count} (Model 2) substantially better than changes in the broader \textbf{\textsc{cyclicArb} transaction count} (Model 1), as indicated by higher adjusted $R^2$ values. This suggests that the chosen market variables are more directly associated with \textsc{cyclicArb} behaviors involving on-chain value extraction (i.e., swaps, \textsc{trade}), rather than speculative probing.

Arbitrum exhibits the best overall model fit, particularly for $\Delta$\texttt{CyclicArbTx} category. This likely reflects a tighter correlation between \textsc{cyclicArb}-\textsc{trade} transaction count and total \textsc{cyclicArb} transaction count on that chain.

\paragraph*{Comparative Analysis of Independent Variables}

\subparagraph*{$\Delta \text{Price}$ (Daily Change in ETH Price).}  
The effect of ETH price fluctuations on \textsc{cyclicArb} activity appears weak and inconsistent across networks:

\begin{itemize}[topsep=2pt, itemsep=1pt, parsep=1pt, partopsep=0pt]
    \item \textbf{Model 1:} No statistically significant association between daily ETH price changes and overall \textsc{cyclicArb} transaction count on any network.
    \item \textbf{Model 2:} A statistically significant negative effect is observed on Base ($-1881.31$, $p < 0.05$), suggesting that a sharp decrease in ETH price may increase the number of \textsc{cyclicArb}-\textsc{trade} executions. The coefficients for Optimism and Arbitrum are also negative but fall short of significance thresholds.
\end{itemize}

These findings suggest a higher prevalence of arbitrage opportunities on days marked by ETH price declines. We propose two primary hypotheses for this observation. Firstly, diminished market liquidity during price downturns \cite{liqchordia2000market} may lead to increased slippage and suboptimal trade execution, thereby fostering greater price discrepancies \cite{adams2024dontletmevslip}. Secondly, the predominance of long leverage in DeFi protocols \cite{heimbach2023defi} could result in a higher frequency of liquidations on days with negative price movements. Such liquidations can trigger significant price swings, which in turn may create further price disparities~\cite{Qin2021}.

\subparagraph*{$\Delta \text{Volatility}$ (Daily Change in ETH Volatility).}  
Volatility, by contrast, plays a more significant role:

\begin{itemize}[topsep=2pt, itemsep=1pt, parsep=1pt, partopsep=0pt]
    \item \textbf{Model 1:} A strong and significant positive effect on Base ($34112.75$, $p < 0.01$) and Arbitrum ($2265.78$, $p < 0.01$), indicating that volatility increases overall \textsc{cyclicArb} transaction activity. No significant effect is observed on Optimism.
    \item \textbf{Model 2:} A similarly strong positive effect is found for \textsc{cyclicArb}-\textsc{trade} transactions on Base ($8919.16$, $p < 0.01$) and Optimism ($1289.53$, $p < 0.01$). On Arbitrum, however, the coefficient is small and not statistically significant.
\end{itemize}

The widely held expectation that increased volatility fuels MEV activity is supported by our results for \textsc{cyclicArb}-\textsc{trade} transactions on Base and Optimism \cite{milionis2023automatedmarketmakingarbitrage, heimbach2024nonatomicarbitragedecentralizedfinance}, as well as for overall \textsc{cyclicArb} counts on Base and Arbitrum. However, the absence of a significant volatility effect on the change of \textsc{cyclicArb}-\textsc{trade} on Arbitrum deserves special attention. In a high-throughput setting with very short block times, an efficient CEX–DEX arbitrage layer can continually realign each DEX price to its CEX counterpart, effectively erasing intra-DEX cyclic arbitrage opportunities, regardless of volatility. Under such a ``hierarchical arbitrage'' regime, volatility may drive cross-venue trades but leave purely on-chain cycles unprofitable. This provides a possible explanation for why we observe little volatility-driven \textsc{cyclicArb} trading on Arbitrum.

\subparagraph*{$\Delta \text{RetailTxs}$ (Change in Retail Trade Count).}  
Retail activity (used here as a proxy for organic user flow) is consistently important for MEV trade behavior:

\begin{itemize}[topsep=2pt, itemsep=1pt, parsep=1pt, partopsep=0pt]
    \item \textbf{Model 1:} Significant positive effects on Base ($23175.65$, $p < 0.05$) and Arbitrum ($16024.00$, $p < 0.01$); not significant on Optimism.
    \item \textbf{Model 2:} Strong and significant on all three networks (i.e., Base, Optimism, and Arbitrum) all at $p < 0.01$.
\end{itemize}

These findings reinforce the idea that user-driven flow is the foundation for MEV extraction via arbitrage.  This is because user transactions, such as trades on DEXs, are the primary actions that perturb market prices. These perturbations create transient price discrepancies across different venues or asset pairs. \textsc{cyclicArb} bots then capitalize on these temporary imbalances by executing arbitrage trades, effectively profiting from the price impact of the initial user-driven activity. Thus, without the initial flow from users, the opportunities for this form of MEV extraction would be significantly diminished. On the other hand, the lack of significance for total \textsc{cyclicArb} transaction count on Optimism suggests that probing or spam-like transactions may be more prevalent there, reducing the signal from genuine trade-driven activity.

\subparagraph*{$\Delta \text{RetailAggFrac}$ (Change in Aggregator Usage by Retail Users).}  
This variable reveals the most striking network divergence:

\begin{itemize}[topsep=2pt, itemsep=1pt, parsep=1pt, partopsep=0pt]
    \item \textbf{Model 1:} Base shows a positive and significant effect ($12230.89$, $p < 0.05$); Arbitrum, a significant negative effect ($-1524.62$, $p < 0.01$); Optimism shows no significance.
    \item \textbf{Model 2:} Base again shows a significant positive effect ($1883.19$, $p < 0.01$), while both Optimism ($-1264.82$, $p < 0.01$) and Arbitrum ($-1319.37$, $p < 0.01$) show significant negative effects.
\end{itemize}

Increased aggregator usage is generally indicative of more efficient transaction routing, which we hypothesized would lead to fewer price discrepancies. However, the Base results present a counterintuitive finding, as higher aggregator utilization did not correspond with a reduction in naive arbitrage opportunities. Several hypotheses may explain this observation:

\begin{itemize}[topsep=2pt, itemsep=1pt, parsep=1pt, partopsep=0pt]
    \item Many tokens on Base may have only one active pool, making direct-to-pool trades effectively MEV-optimal (i.e., they do not open up an arbitrage opportunity) --- undermining the aggregator efficiency signal.
    \item Aggregator usage may correlate with increased activity in major token pairs (e.g., ETH-stables), especially during volatile periods, making \texttt{RetailAggFrac} a latent proxy for volatility.
\end{itemize}

Thus, our use of \texttt{RetailAggFrac} as a proxy for the ratio between \text{MEV-optimal Trades} and  \text{Total Trades}, where MEV-optimal trades are those that do not introduce new arbitrage opportunities, may be misaligned on Base due to its unique market structure.

On Optimism and Arbitrum, the expected pattern emerges: increased aggregator usage correlates with reduced \textsc{cyclicArb}-\textsc{trade} activity, consistent with the role of aggregators in mitigating simple arbitrage opportunities by improving trade routing.

\paragraph*{Summary of Insights}

The regression results demonstrate that some drivers of \textsc{cyclicArb} (such as user trade count) are robust across networks. In contrast, other factors like volatility and aggregator usage show strong network-specific effects, reflecting differences in market structure and protocol design. These findings underscore that \textsc{cyclicArb} dynamics are not uniform across Layer 2s, but are shaped by their unique configurations, including block times, mempool behavior, and execution environments.

\section{Discussion}

Optimistic MEV fits into the broader MEV landscape as a new variant of an old problem that was thought to have been largely solved. Its optimistic execution model is novel because profitability is determined on-chain during transaction execution, but the externality it creates in the form of large volumes of low-value or failed transactions mirrors and even outpaces the spam problem that plagued early Ethereum MEV. On Ethereum, that earlier spam arose from bots flooding the mempool with precomputed transactions and engaging in priority gas auctions without knowing whether their transactions would land in the right place in the block to secure the opportunity, a dynamic that private orderflow and MEV auctions were introduced to mitigate.

This behavior has significant consequences. It generates large volumes of optimistic MEV transactions on Layer 2s, creating spam that consumes blockspace without delivering corresponding value to users. This spam constitutes a negative externality for the network and must still be stored by nodes. The resulting data burden increases costs for infrastructure providers and risks concentrating node operation among better-resourced actors. It also appears that optimistic MEV creates a workload that very often fills up blocks, which should be taken into account when designing blockchains.%

For searchers, this environment reshapes the underlying economics. The focus shifts away from high-certainty execution toward balancing the cost of repeated speculative probes compared to the profit they generate. Capturing value becomes less of a latency race; instead, the transactions are submitted before the opportunities even exist. Further, optimistic MEV requires paying far lower priority fees compared to traditional MEV on Ethereum Layer 1, which fundamentally changes how bots compete and extract value.

MEV auctions are the main mechanism used in practice to prevent the negative externalities of MEV by reducing spam. Optimistic MEV makes this harder to apply. Its speculative, high-frequency nature could produce a flood of bids and overwhelm an auction with noise, while the short block intervals on Layer 2s leave little room to run the process at all. Revert protection, as implemented by Unichain, offers an alternative mitigation but risks generating a similar kind of spam directed at the sequencer. In addition, the presence of centralized sequencers on Layer 2s creates incentive dynamics very different from Layer 1. 

Addressing optimistic MEV will likely require rethinking how, or whether, MEV auctions and related mechanisms can work in rollup environments. Our results suggest that network design choices can influence the prevalence of optimistic MEV: the significantly lower rate of optimistic MEV we observe on Arbitrum compared to OP-stack rollups indicates that factors such as gas pricing or block intervals can meaningfully affect these strategies. This points toward protocol-level levers for reducing optimistic MEV as Layer 2 designs evolve.

\section{Outlook and Conclusion}
In this work, we investigate the significant and large-scale impact of \textit{optimistic MEV} on Layer 2 networks, focusing on cyclic arbitrage, which on these networks is predominantly executed in this optimistic form and accounts for more than 50\% of gas usage in OP-Stack ecosystems. Our findings reveal that a confluence of factors, namely low transaction fees, extended interblock times, and PGA ordering mechanisms prevalent on OP-Stack chains, contributes to this extensive MEV activity.

The ramifications of this activity are substantial, manifesting as a phenomenon akin to network spam. This spam-like behavior inundates the network with low-value transactions, leading to inefficient resource allocation, wasted chain space, and potential degradation of user experience. Critically, this also limits the network’s capacity to scale for higher-value or user-driven activity, as blockspace is increasingly consumed by speculative probing. 

Addressing this challenge is essential for the long-term sustainability of these networks. A shift in network dynamics is necessary, either through the emergence of organic demand willing to pay higher fees, thereby crowding out low-value MEV attempts, or through direct interventions such as raising minimum transaction costs. Arbitrum, where different design choices have demonstrated some efficacy in curbing similar behaviors, serves as a pertinent case study for the latter approach.

Further differentiating the network dynamics, our analysis uncovered that most unsuccessful MEV attempts do not revert on OP-Stack Layer 2s, a stark contrast to Arbitrum, where such reversions are common. This divergence is likely attributable to Arbitrum's FCFS transaction ordering, approximately tenfold shorter interblock times, and a distinct fee market. These elements collectively appear to incentivize more deterministic behavior from MEV bots on Arbitrum, as evidenced when comparing the strategies of top bots across platforms, unlike the more speculative attempts observed on Optimism and Base. This underscores how architectural and fee-market designs can significantly influence the strategies of MEV bots and overall network congestion.

\bibliography{lipics-v2021-sample-article}
\newpage
\appendix

\section{Contract Similarity}\label{app:clone}
For every address\footnote{Although we focus only on the primary contracts, many arbitrage transactions also invoke auxiliary, non-DEX ``helper'' contracts during execution.} that survives the manual audit, we retrieve the bytecode directly from the nodes using the JSON–RPC call \texttt{eth\_getCode(\,<address>,\ "latest")}. Next, we use the Heimdall disassembler~\cite{Becker2025Heimdall} to convert this bytecode into a sequence of EVM opcodes. Following the method used by Kiffer et al.~\cite{10.1145/3278532.3278575}, we remove all operand data so that only the opcode mnemonics remain. We then slide a five‐opcode window over the cleaned stream and count the occurrences of each unique chunk, producing a high-dimensional frequency vector for each contract. Finally, we compute the cosine similarity between any two vectors to quantify how closely their opcode patterns match. The key steps are summarized in the following pseudocode:

\begin{algorithm}
\caption{Compute Contract Similarity Between Two Contracts}
  \label{nearestGridPoint}
  \begin{algorithmic}[1]
    \Function{Similarity}{bytecode$_1$, bytecode$_2$}
      \State dis$_1$, dis$_2$ $\gets$ disassemble(bytecode$_1$), disassemble(bytecode$_2$)
      \State opcs$_1$, opcs$_2$ $\gets$ strip\_opcodes(dis$_1$), strip\_opcodes(dis$_2$)
      \State freq\_vec$_1$,freq\_vec$_2$ $\gets$ compute\_freq\_vecs(c$_1$=opcs$_1$, c$_2$=opcs$_2$, N=5)
      \State \Return $\frac{\text{freq\_vec$_1$} \cdot \text{freq\_vec$_2$}}{\|\text{freq\_vec$_1$}\|  \|\text{freq\_vec$_2$}\|}$
    \EndFunction
  \end{algorithmic}
\end{algorithm}

\begin{figure}[htb]
    \centering
    \includegraphics[width=0.8\linewidth]{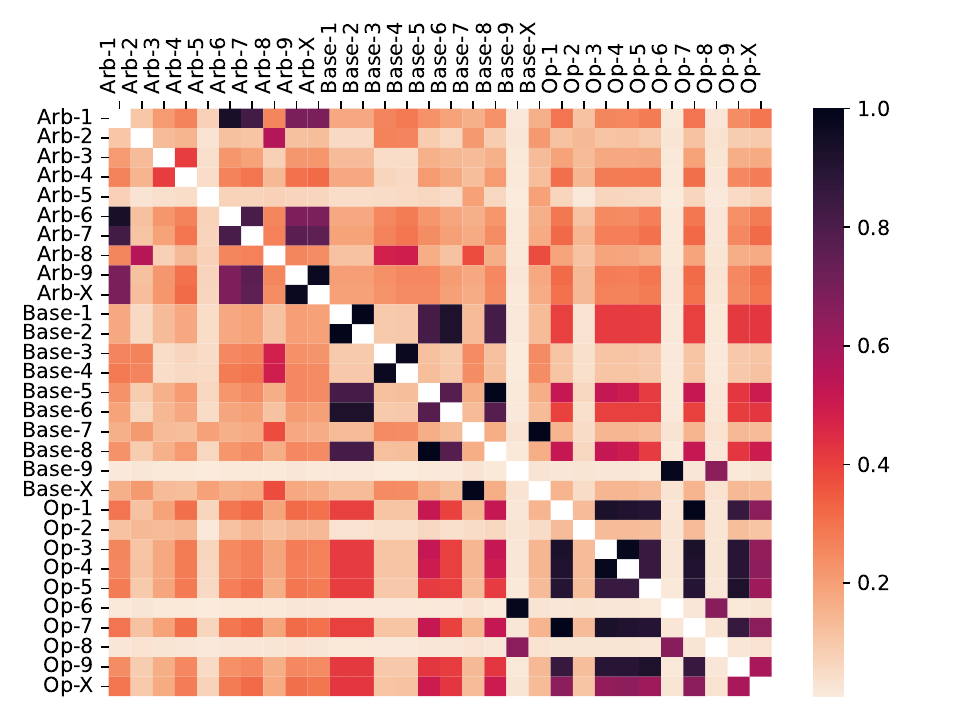}
    \caption{Cosine similarity scores between MEV bot contracts on Arbitrum, Base, and Optimism. Each block along the diagonal (e.g., Arbitrum-1 through Arbitrum-X, Base-1 through Base-X, Op-1 through Op-X) shows intra-network similarity, while off-diagonal blocks reveal inter-network code reuse. High similarity values (closer to 1.0) indicate shared or cloned implementations, with particularly tight clusters visible among Base contracts.}
    \label{fig:similarity_all}
\end{figure}

Figure~\ref{fig:similarity_all} visualizes the pairwise cosine similarity between MEV bot contracts on Arbitrum, Base, and Optimism using 5-opcode frequency vectors derived from each contract’s disassembled bytecode. Along the diagonal, intra-network comparisons reveal that Base contracts form tight clusters, indicating many bots share nearly identical implementations, while Optimism shows a moderate level of code reuse and Arbitrum exhibits the greatest internal diversity. Off-diagonal blocks show inter-network similarity: Base and Optimism share some common code patterns, likely reflecting shared bot frameworks, but Arbitrum contracts remain largely distinct from those on the other two chains. The prominent high-similarity bands on Base confirm widespread cloning or redeployment of identical bot logic, whereas the lighter, more sporadic similarities elsewhere suggest more heterogeneous or independently developed MEV implementations.

\section{Reverts}
\label{revertsapp}
Contrary to common assumptions~\cite{miller2024l2congestion}, we find that most transaction reverts on Base are not driven by failed cyclic arbitrage attempts but by event-driven``liquidity-sniping'' strategies \cite{sniper,colkitt2024sniper}. In these cases, MEV bots monitor the on-chain pool for new token listings and submit purchase transactions in the same block that liquidity is added. The first significant surge in revert rates on Base coincided with bots back-running FriendTech share listings as soon as new accounts launched~\cite{friend-tech}. Although Base’s mempool is private, a transient transaction-pool leak enabled MEV bots to execute same-block back-runs~\cite{bantg2023xpost} until the vulnerability was patched~\cite{opgeth-pr118}. MEV bots exploited the sequencer’s ordering by matching user gas bids and spamming identical transactions; any duplicate that landed before the user’s own transaction reverted, yet the low gas prices on Base made this spray-and-pray approach profitable, as illustrated in block \href{https://basescan.org/block/2930614}{2930614}~\cite{bertcmiller-mev-base,bertcmillerspamlast2023}.

\begin{figure}[htb]\vspace{-10pt}
    \centering
    \includegraphics[width=\linewidth]{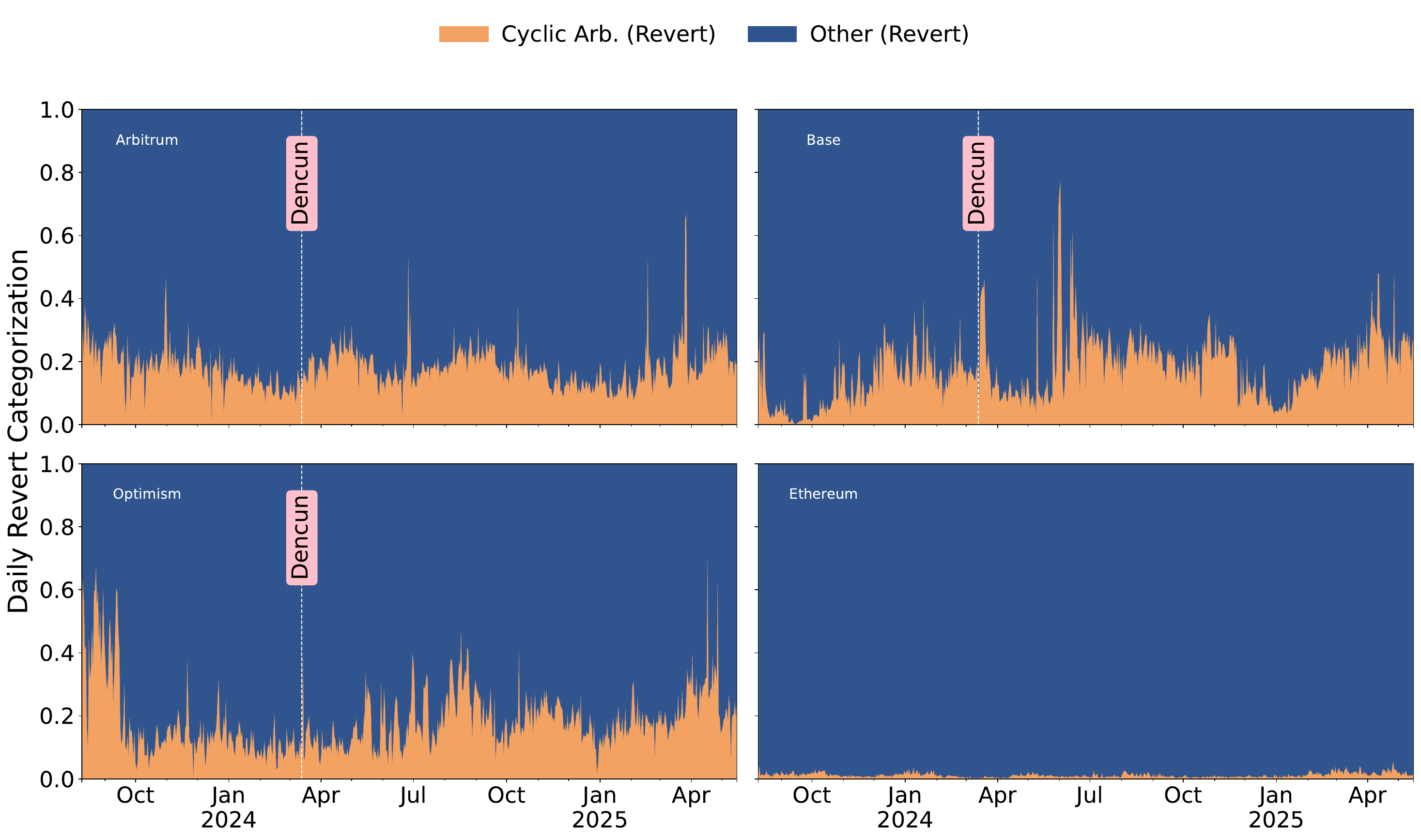}
    \caption{Daily share of reverted transactions on Arbitrum, Base, Optimism, and Ethereum Layer 1, split between cyclic arbitrage MEV bots (``Cyclic Arb. (Revert)'') and all other activity (``Other (Revert)'').}
    \label{fig:reverts}\vspace{-10pt}
\end{figure}

Figure~\ref{fig:reverts} shows that, although optimistic-MEV bots account for a large share of total gas usage on Base and Optimism, they represent a substantially smaller fraction of reverted transactions. This disparity indicates that most reverts on these networks stem from non-optimistic-MEV activity, chiefly liquidity-sniping, rather than failed arbitrage attempts.

Additionally, around late November 2024 and early January 2025, we see the most increase, in absolute terms, in reverts on Base, which causes network fees to increase. These are once again caused by sniper bots interacting with applications such as Clanker \cite{clanker,colkitt2024sniper} and Virtuals \cite{virtuals}. Importantly, we note that these drive the median gas price paid by MEV bots in our dataset during those periods as can be seen in the Figure~\ref{fig:gas_price}.

\section{Data Collection}\label{app:datacollection}
We collect data for Base~\cite{base2025}, Optimism~\cite{optimism2025}, Arbitrum~\cite{arbitrum2025} and Ethereum. At the time of writing, Base, Optimism, and Arbitrum are the three largest Layer 2s based on  \emph{total value locked (TVL)} according to L2BEAT~\cite{l2beat2025,growthepie2025}. For comparison, we collect the same data for Ethereum too, enabling us to characterize the differences between Layer 1 and Layer 2,  especially regarding the prevalence of optimistic MEV.

Our empirical analysis relies on a comprehensive dataset primarily sourced from Dune Analytics~\cite{Dune2025DataFreshness}, a platform offering comprehensive access to raw and decoded blockchain data across various networks, including all networks of interest for this study. We collect the respective data spanning from August 2023 to May 2025. This dataset encompasses several critical types of data sources: 
\begin{enumerate}[topsep=2pt, itemsep=1pt, parsep=1pt, partopsep=0pt]
    \item \textbf{Transactional Data:} We retrieve high-level transaction details including, but not limited to, sender and recipient addresses, gas utilized, transaction value, input data, block numbers, and timestamps. Crucially, we also collect detailed transaction traces, which include internal transactions, emitted logs (events), and relevant state changes. This granular data was primarily sourced by querying Dune Analytics' tables such as \texttt{\{\{chains\}\}.transactions} (e.g. \cite{dune2025transactions}) and \texttt{\{\{chains\}\}.traces} (e.g. \cite{dune2025traces}) for Ethereum Layer 1 and each respective Layer 2 network.
    \item \textbf{Decentralized Exchange Activity Data:} To analyze MEV transactions, we collect extensive data on trades and liquidity pools from DEX protocols operating on the target Layer 2s. This includes detailed swap event data and liquidity pool contract addresses, leveraging curated datasets such as those described in~\cite{dune2025trades}.
    \item \textbf{Contract Address Identification:} A curated list of smart contract addresses relevant to DEX activities, such as prominent DEX routers and aggregator contracts on the studied Layer 2s, is compiled. This is achieved through a combination of leveraging existing tagged address lists within Dune Analytics \cite{Dune2025DataFreshness}, cross-referencing them with on-chain explorers \cite{arbiscan,optimistic,basescan,intelarkm}, and preliminary heuristic-based identification from our dataset.
    \item \textbf{Contract Bytecodes:} For the contracts that we marked on Layer 2s, we fetch their bytecodes to be utilized in future steps to measure contract similarity (see \cref{app:clone}). We run our own archive nodes for Base, Arbitrum, and Optimism networks to fetch the bytecodes using \texttt{JSON-RPC} method \texttt{eth\_getCode}. 
    \item \textbf{Price Data:} We additionally obtain daily Open-High-Low-Close (OHLC) data for the ETH price on Ethereum Layer 1 using Dune Analytics \cite{Dune2025DataFreshness}, which we use for regression analysis and volatility calculation.
\end{enumerate}
This multi-faceted data collection strategy was designed to provide the necessary inputs for our MEV and transaction classification pipeline, enabling a robust identification and empirical analysis of various MEV activities on the selected Layer 2 networks. We acknowledge Dune Analytics' data ingestion processes, including considerations for data freshness (up to a day of delay) and the support of all major DEXs~\cite{dune2025trades}. The few missing DEXs (such as those involving 1inch OTC trades) do not impact our results, as they are not relevant to MEV activity. 
\section{Cyclic Arbitrage Contract Detection (Extended)}
\label{app:classification}

This section explains the full cyclic arbitrage contract detection process shown in \cref{sec:classification:stage1} in more detail.
This classification stage aimes to identify smart contracts exhibiting strong on-chain signs consistent with cyclic arbitrage activities. 

The classifier was implemented entirely in Dune SQL, leveraging Dune Analytics' data tables and Torino engine. The core tables utilized are described in the following way:
\begin{description}
    \item[dex\_aggregator.trades \cite{duneaggt2025}] Utilized to identify and subsequently filter out known DEX aggregator and router contract addresses, as these primarily act as intermediaries rather than originating MEV actors, thereby reducing noise. 
    \item[dex.trades \cite{dune2025trades}] Provided swap events from DEXs, including essential fields such as \texttt{tx\_hash}, event index within the transaction (\texttt{evt\_index}), transacted tokens, and amounts involved in each swap.
    \item[dex.addresses \cite{dune2025labels}] Served as a label database, crucial for identifying contract types (e.g., routers, liquidity pools, factories) and distinguishing them from potential MEV bot contracts.
\end{description}
\noindent
This process is formalized as follows:\\\\ \noindent Let $\mathbb A$ be the set of all \emph{externally owned accounts (EOAs)}, $\mathbb{K}$ the set of contract addresses, a \textbf{swap event}, denoted as $s$, is defined as a 5-tuple:
$$s = (\text{token}_{\text{sold}}, \text{token}_{\text{bought}}, \text{amount}_{\text{sold}}, \text{amount}_{\text{bought}}, \text{idx}) \in \mathbb{K} \times \mathbb{K} \times \mathbb{R}^+ \times \mathbb{R}^+ \times \mathbb{N}$$
where:
\begin{itemize}[topsep=2pt, itemsep=1pt, parsep=1pt, partopsep=0pt]
    \item $\text{token}_{\text{sold}}$ is the contract address of the token sold.
    \item $\text{token}_{\text{bought}}$ is the contract address of the token bought.
    \item $\text{amount}_{\text{sold}}$ is the quantity of $\text{token}_{\text{sold}}$.
    \item $\text{amount}_{\text{bought}}$ is the quantity of $\text{token}_{\text{bought}}$.
    \item $\text{idx}$ is the intra-transaction index of the emitted swap log, preserving execution order.
\end{itemize}

Let $\mathcal{S}$ be the set of all possible swap events. A \textbf{trade transaction}, denoted as $tr$, associated with a single blockchain transaction, is the set of all swap events $s \in \mathcal{S}$ that occurred within that transaction. Thus, $tr \in \mathcal P \mathcal (S)$. \\\\ \noindent A \textbf{transaction} is represented by a 4-tuple: $$(\text{hash}, \text{to}, \text{from}, tr),$$
where
\begin{itemize}[topsep=2pt, itemsep=1pt, parsep=1pt, partopsep=0pt]
    \item $\text{hash}$ is the transaction hash.
    \item $\text{to} \in \mathbb{K}$ is the contract address that the transaction invoked first.
    \item $\text{from} \in \mathbb{A}$ is the EOA that initiated the transaction.
    \item $tr$ is the trade (set of swap events) executed within this transaction.
\end{itemize}
Let $\mathcal{T}_{\text{raw}}$ be the set of all such transactions containing at least one swap event. We define a feature extraction function $f: \mathcal{P}(\mathcal{S}) \to  \text{Seq}(\mathbb{K}) \times \text{Map}(\mathbb{K}, \mathbb{R}) $, where $\text{Seq}(\mathbb{K})$ is the space of sequences of token addresses and $\text{Map}(\mathbb{K}, \mathbb{R})$ is the space of mappings from token addresses to real-valued balance changes.
For a given trade $tr = \{s_1, \dots, s_k\}$, these swaps are first ordered by their $\text{idx}$ value, yielding an ordered sequence $\tilde{tr} = \langle \acute{s}_1, \acute{s}_2, \dots, \acute{s}_k \rangle$, where $\acute{s}_j = (\text{token}_{\text{sold},j}, \text{token}_{\text{bought},j})$.
The function $f(tr)$ then produces a pair $(\Pi, \Delta_B)$:
\begin{enumerate}[topsep=2pt, itemsep=1pt, parsep=1pt, partopsep=0pt]
    \item $\Pi = \langle \text{token}_{\text{sold},1}, \text{token}_{\text{bought},1},
    \dots, \text{token}_{\text{sold},k}, \text{token}_{\text{bought},k} \rangle$. This is an ordered sequence of token addresses reflecting the actual path of token conversions in the transaction.
    \item $\Delta_B = \{ (\kappa, \delta_\kappa) | \kappa \in \mathbb{K}, \delta_\kappa \in \mathbb R \}$. This is a map representing the net balance changes after the transaction from the initiator's perspective across all tokens involved in the trade $tr$. Specifically, for each token $\kappa$, $\delta_\kappa = \sum_{j=1}^k (\text{amount}_{\text{bought},j} \cdot \mathbb{I}(\text{token}_{\text{bought},j} = \kappa) - \text{amount}_{\text{sold},j} \cdot \mathbb{I}(\text{token}_{\text{sold},j} = \kappa))$, where $\mathbb{I}(\cdot)$ is the indicator function.
\end{enumerate}
Let $\mathcal{T}'$ be the set of transactions transformed by $f$:
$$\mathcal{T}' = \{ (\text{hash}, \text{to}, \text{from}, f(tr)) \mid (\text{hash}, \text{to}, \text{from}, tr) \in \mathcal{T}_{\text{raw}} \}$$
An element in $\mathcal{T}'$ is of the form $(\text{hash}, \text{to}, \text{from}, (\Pi, \Delta_B))$. Then, we apply three sequential filters ($\alpha_1, \alpha_2, \alpha_3$) to $\mathcal{T}'$:
\begin{enumerate}[topsep=2pt, itemsep=1pt, parsep=1pt, partopsep=0pt]
    \item \textbf{Filter $\alpha_1$ (Router/Aggregator Exclusion):}
    Let $\mathcal{R}_{\text{contracts}}$ and $\mathcal{A}_{\text{contracts}}$ be the sets of known router and aggregator contract addresses, respectively. This filter removes transactions directly interacting with these intermediary contracts.
    $$\mathcal{T}^{(1)} = \alpha_1(\mathcal{T}') = \{ (\text{hash}, \text{to}, \text{from}, (\Pi, \Delta_B)) \in \mathcal{T}' \mid \text{to} \notin (\mathcal{R}_{\text{contracts}} \cup \mathcal{A}_{\text{contracts}}) \}$$

    \item \textbf{Filter $\alpha_2$ (Cyclic Swap Detection):}
    This filter identifies transactions whose sequence of token swaps
    $\Pi = \langle \pi_1, \pi_2, \dots, \pi_{2k} \rangle$ forms a cycle. The predicate $\text{isCyclic}(\Pi)$ holds true if:
    \begin{itemize}[topsep=2pt, itemsep=1pt, parsep=1pt, partopsep=0pt]
        \item $k \ge 1$ (i.e., there is at least one swap).
        \item $\pi_1 = \pi_{2k}$ (the first token sold is the same as the last token bought).
        \item For all $j \in \{1, \dots, k-1\}$, $\pi_{2j} = \pi_{2j+1}$ (the token bought in the $j$-th swap is the same as the token sold in the $(j+1)$-th swap, ensuring path continuity).
    \end{itemize}
    $$\mathcal{T}^{(2)} = \alpha_2(\mathcal{T}^{(1)}) = \{ (\text{hash}, \text{to}, \text{from}, (\Pi, \Delta_B)) \in \mathcal{T}^{(1)} \mid \text{isCyclic}(\Pi) \}$$

    \item \textbf{Filter $\alpha_3$ (Profitability Assessment):}
    This filter retains transactions that result in a net profit for a token without incurring losses in any other token. It evaluates the balance changes map $\Delta_B$. The predicate $\text{isProfitable}(\Delta_B)$ holds true if:
    \begin{itemize}[topsep=2pt, itemsep=1pt, parsep=1pt, partopsep=0pt]
        \item There exists at least one token $\kappa$ such that its balance change $\delta_\kappa$ is strictly positive ($\delta_\kappa > 0$).
        \item For all token balance changes $\delta_{\kappa'}$, $\delta_{\kappa'} \ge 0$ (i.e., there are no negative balance changes).
    \end{itemize}
    $$\mathcal{T}^{(3)} = \alpha_3(\mathcal{T}^{(2)}) = \{ (\text{hash}, \text{to}, \text{from}, (\Pi, \Delta_B)) \in \mathcal{T}^{(2)} \mid \text{isProfitable}(\Delta_B) \}$$
\end{enumerate}
Combining these filters, the set of transactions identified as high-probability cyclic arbitrage is $\mathcal{T}_{\text{final}} = \mathcal{T}^{(3)}$.
As the final step of the algorithmic pre-filter, the set of contract addresses marked as high-probability cyclic arbitrage bots, $\mathcal{C}_{\text{bot}}$, is derived from these transactions:
$$\mathcal{C}_{\text{bot}} = \{ \text{to} \mid (\text{hash}, \text{to}, \text{from}, (\Pi, \Delta_B)) \in \mathcal{T}_{\text{final}} \}$$
These contracts are then passed to the subsequent validation stage.

\section{Cyclic Arbitrage Contract Validation}
\label{app:manaudit}
To verify that the data set does not contain false positives, we inspect the candidate contracts. We perform these validation steps for each chain and the set of respective candidate contracts until more than 80\% of all the gas used in transactions involving these contracts originates from validated contracts.
Our validation process consists of the following four steps:

\begin{itemize}[topsep=2pt, itemsep=1pt, parsep=1pt, partopsep=0pt]
    \item \textbf{Code Verification and Labeling:} We investigate whether the contract's source code was verified on public block explorers (e.g., Arbiscan and its forks \cite{unichain,arbiscan,optimistic,basescan}). Verified contracts that were not proxies and had clear, non-MEV related functionalities, such as standard token contracts or well-known application logic, were excluded. We further query Arkham~\cite{intelarkm} and community-curated datasets \cite{EntropyDune2025,libmev}, and exclude contracts with labels that are inconsistent with MEV-bot activity.
    \item \textbf{Transaction Trace Analysis for Non-Trading Activity:} For transactions initiated by a candidate contract that did not result in direct swaps or token transfers, we inspect their execution traces to see if they interacted with DEX-related contracts (e.g., calling functions such as \texttt{getReserves} or \texttt{slot0} on DEXs to read pool reserves, or interacting with periphery contracts).
    We exclude all candidate contracts that do not have a clear majority (60\% and more) of such DEX-related interaction. Note that the few candidates removed all had very low interaction numbers (between 0\% and 30\%), while the large majority of candidates have a very high DEX-interaction rate (85\% or more).
    \item \textbf{Caller Profile Analysis:} We examined the diversity of callers interacting with the candidate contract and their frequency. If a candidate contract had (i) more than 3 distinct, and unrelated EOAs interacting with it and (ii) the frequency of interaction could credibly have been human, we consider the contract to not belong to a cyclic arbitrage bot. With this validation heuristic, we err on the side of caution, as there are cases where specialized MEV bots are triggered by multiple EOAs.
    \item  \textbf{Swaps per Transaction:} We analyze the distribution of swaps per transaction (of those transactions that involve at least one swap) by observing the 10th, 25th, 50th, 75th, and 90th percentiles. A 10th or 25th percentile value of one ($P_{10} = 1$ or $P_{25}=1$) would serve as a strong indicator that the contract does not belong to a cyclic arbitrage bot, as a cyclic arbitrage requires at least two transactions. For $P_{10},P_{25},P_{50},P_{75},P_{90}$ values of the top 10 contracts on each analyzed Layer 2, consult the Table ~\ref{tab:mev_cycle}. All candidate contracts pass this validation stage.
\end{itemize}

The few misclassified contracts we find during validation are mainly due to popular mislabeled routers or public utility contracts. The full list of removed contracts is found in \cite{non-anonymous2025optimistic}. These contracts are removed and the remaining contracts form our final set of cyclic arbitrage contracts $\mathcal{C}_{\text{bot}}$. 

\begin{table}[t!]
\begin{subtable}[h]{\textwidth}
  \centering
  \setlength{\tabcolsep}{3pt} %

  \centering

  \setlength{\tabcolsep}{3pt} %
  \sisetup{
  }
  \begin{tabular*}{0.95\linewidth}{@{\extracolsep{\fill}} 
    l %
    S[table-format=9, group-separator={,}]  %
    S[table-format=9, group-separator={,}]  %
    S[table-format=9, group-separator={,}]  %
    S[table-format=10, group-separator={,}]                      %
    S[table-format=10] %
    @{}
  }
    \toprule
    {contract} & %
    {$P_{10}$} &    %
    {$P_{25}$} &   %
    {$P_{50}$} &  %
    {$P_{75}$} &  %
    {$P_{90}$} \\ %
    \midrule
    \href{https://basescan.org/address/0xf5ff765b0c1278e54281193d7019281e0e50a8c0}{\texttt{0xf5ff\dots a8c0}} & 2 &  2 & 3 &    3 &  5 \\
    \href{https://basescan.org/address/0xbff69c6f5bb902807b05d0d1a523e3b20b3b73e9}{\texttt{0xbff6\dots 73e9}} & 2 & 2 & 2 & 3 & 4 \\
    \href{https://basescan.org/address/0xaa877b9d9fbd4d6c40c3b9503f0d251531f1fb84}{\texttt{0xaa87\dots fb84}} & 2 &  2 & 2 &  2 & 2 \\
    \href{https://basescan.org/address/0xddfa17794fdd2903f16a496ebefb76a87355abd3}{\texttt{0xddfa\dots abd3}}  &  2 &  2 &  2 &   3 & 4 \\
    \href{https://basescan.org/address/0xdade45a1acda95c981040aaedb6fe30cf3bc6084}{\texttt{0xdade\dots 6084}} & 2 &  2 &  2 &   3 & 5 \\
    \href{https://basescan.org/address/0xe91ca2ee0566e76c99321435e9df5817b9c996af}{\texttt{0xe91c\dots 96af}} &  2 &  2 &  3 &3 & 6 \\
    \href{https://basescan.org/address/0x4e85efbb92d53d1ce0e8a0e5129b881c5ffa0cea}{\texttt{0x4e85\dots 0cea}} & 2 &  2 &  3 &   3 & 5 \\
    \href{https://basescan.org/address/0x2b24051780aeea28aec159ae8e46528fafe5446f}{\texttt{0x2b24\dots 446f}}  &  2 &  2 &  3 &    3 & 5 \\
    \href{https://basescan.org/address/0x826fd727477547bd89d75f7941d35f525c04b5f5}{\texttt{0x826f\dots b5f5}} & 2 & 2 &  3 &    3 & 6 \\
    \href{https://basescan.org/address/0xbba96510886d6abbb917da35add5362cd8e33cf9}{\texttt{0xbba9\dots 3cf9}} &  2 &  2 &  3 &   3 & 3 \\
    \bottomrule
  \end{tabular*}\vspace{3pt}
  
  \caption{Base}
  \label{tab:cycle_base}
\end{subtable}

\begin{subtable}[h]{\textwidth}
  \centering

  \setlength{\tabcolsep}{3pt} %
  \sisetup{
  }
  \begin{tabular*}{0.95\linewidth}{@{\extracolsep{\fill}} 
    l %
    S[table-format=9, group-separator={,}]  %
    S[table-format=9, group-separator={,}]  %
    S[table-format=9, group-separator={,}]  %
    S[table-format=10, group-separator={,}]                      %
    S[table-format=10] %
    @{}
  }
    \toprule
    {contract} & %
    {$P_{10}$} &    %
    {$P_{25}$} &   %
    {$P_{50}$} &  %
    {$P_{75}$} &  %
    {$P_{90}$} \\ %
    \midrule
\href{https://optimistic.etherscan.io/address/0x887290c34856cd3ba1b84da78cccf43812f66324}{\texttt{0x8872\dots 6324}} & 2& 2 & 3 & 4 & 8\\
\href{https://optimistic.etherscan.io/address/0xabf4daac18925530d1e4f99fd538d57b8bf1017c}{\texttt{0xabf4\dots 017c}}& 2 & 2 & 2 & 3 & 6\\
\href{https://optimistic.etherscan.io/address/0xcdcca22e3b1e0170532eea4b14e5fb3a73b45c8f}{\texttt{0xcdcc\dots 5c8f}} & 2 & 3 & 3 & 3 & 6\\
\href{https://optimistic.etherscan.io/address/0xd3dc079ac6f98275bbbcd0aff11cbaadb4d8f2ac}{\texttt{0xd3dc\dots f2ac}} & 2 & 2 & 3 & 3 & 6 \\
\href{https://optimistic.etherscan.io/address/0x4d43aa8a0d161a7f2e7ba6c4226fe67998ee987f}{\texttt{0x4d43\dots 987f}} & 2 & 2 & 3 & 4 & 7\\ 
\href{https://optimistic.etherscan.io/address/0x3955945c12eb164d2356de1a290ec54fa4c574c2}{\texttt{0x3955\dots 74c2}} & 2 & 2 & 3 & 3 & 6 \\ 
\href{https://optimistic.etherscan.io/address/0x9d1b033ac8bff2b07fb7d13385b8c270db25f96f}{\texttt{0x9d1b\dots f96f}} & 2 & 2 & 3 & 4 & 7\\ 
\href{https://optimistic.etherscan.io/address/0x0daf895a78eb49151a5f1003818939770d3ca7dd}{\texttt{0x0daf\dots a7dd}} & 2 & 2 & 3 & 3 & 6\\
\href{https://optimistic.etherscan.io/address/0x2642f2b26cce1b96f2a94ee1660d5e7c7b9839e6}{\texttt{0x2642\dots 39e6}} & 2 & 2 & 2 & 3 & 4\\
\href{https://optimistic.etherscan.io/address/0xf2610cca617c94d3b617ec5872d191156bcfc883}{\texttt{0xf261\dots c883}} & 2 & 2 & 3 & 5 & 9\\
    \bottomrule
  \end{tabular*}\vspace{3pt} 
  
  \caption{Optimism}
  \label{tab:cycle_op}
\end{subtable}

\begin{subtable}[h]{\textwidth}
  \centering
  \setlength{\tabcolsep}{5pt} %
  \sisetup{
  }

  \begin{tabular*}{0.95\linewidth}{@{\extracolsep{\fill}} 
    l %
    S[table-format=9, group-separator={,}]  %
    S[table-format=9, group-separator={,}]  %
    S[table-format=9, group-separator={,}]  %
    S[table-format=10, group-separator={,}]                      %
    S[table-format=10] %
    @{}
  }
    \toprule
    {contract} & %
    {$P_{10}$} &    %
    {$P_{25}$} &   %
    {$P_{50}$} &  %
    {$P_{75}$} &  %
    {$P_{90}$} \\ %
    \midrule
\href{https://arbiscan.io/address/0x00000000cfe3369bcdbc76071ba6e0a4e0fe98bd}{\texttt{0x0000\dots 98bd}} & 2 & 2 & 3 & 3 & 4\\
\href{https://arbiscan.io/address/0x60ca5def792e7528f8ff48b7aea60c1e2234294b}{\texttt{0x60ca\dots 294b}} & 2 & 3 & 3 & 3 & 3 \\
\href{https://arbiscan.io/address/0x689368aae04fe7d961d1e545b28266c4f179151c}{\texttt{0x6893\dots 151c}} & 2 & 2 & 3& 3 & 4\\
\href{https://arbiscan.io/address/0x9e52272477c2fcbe2c38bf12dd06cb5739975867}{\texttt{0x9e52\dots 5867}} & 2 & 2 & 3 & 3 & 4\\
\href{https://arbiscan.io/address/0xa9ff271ee217dc1c9ce3f7ebf0d6f096842cd82f}{\texttt{0xa9ff\dots d82f}} & 2 & 2 & 2 & 2 & 3\\
\href{https://arbiscan.io/address/0x00000000c868e634590a981dd3bd007dca5c2e43}{\texttt{0x0000\dots 2e43}} & 2 & 2 & 3 & 3 & 4\\
\href{https://arbiscan.io/address/0x0000b0ca00000017d543c800be4257629544fb51}{\texttt{0x0000\dots fb51}} & 2 & 2 & 3 & 3 & 4 \\
\href{https://arbiscan.io/address/0xf238d4353246948ecdc3f3252ae939fe5234e145}{\texttt{0xf238\dots e145}} &  2 & 2 & 3 & 3 & 3 \\ 
\href{https://arbiscan.io/address/0xe98b3b8d43842a52efc848bf10be28d65c2ed87c}{\texttt{0xe98b\dots d87c}} & 2 & 2 & 2 & 2 & 2 \\
\href{https://arbiscan.io/address/0x84f1d597f1160918d9ec28c5ab0b2eb5394bb7cf}{\texttt{0x84f1\dots b7cf}} & 2 & 2 & 2 & 2 & 2\\
    \bottomrule
  \end{tabular*}\vspace{3pt}
  
  \caption{Arbitrum}
  \label{tab:cycle_arb} 
\end{subtable}
\caption{This table reports statistics for the top 10 MEV bot contracts (ranked by cumulative gas usage) on Base (Table~\ref{tab:cycle_base}), Optimism (Table~\ref{tab:cycle_op}), and Arbitrum (Table~\ref{tab:cycle_arb}). For each contract, we report 10th, 25th, 50th, 75th and 90th percentile of the number of swaps in a transaction which include at least one swap. The query associated with the tables can be viewed at \cite{dunequery5534551}.} \label{tab:mev_cycle}
\end{table}

\section{Arbitrum: Other (Interaction)}\label{app:arbitrum}
The relative plot in Figure~\ref{fig:gas-norm} reveals that the \textsc{other} - \textsc{interaction} category accounts for a significantly larger share of gas consumption on \textbf{Arbitrum} compared to other chains. An analysis of these transactions identified the primary gas-consuming contracts. Uniswap's \textbf{V3 Position Manager} (\href{https://arbiscan.io/address/0xC36442b4a4522E871399CD717aBDD847Ab11FE88}{\texttt{0xC364...FE88}}) was the single largest consumer, responsible for nearly \textbf{10\%} of the gas in this category. It was followed by a variety of other contracts, including Uniswap's V3 Routers (\href{https://arbiscan.io/address/0xe592427a0aece92de3edee1f18e0157c05861564}{\texttt{0xe592...1564}}, \href{https://arbiscan.io/address/0x68b3465833fb72a70ecdf485e0e4c7bd8665fc45}{\texttt{0x68b3...fc45}}), contracts associated with on-chain games (\href{https://arbiscan.io/address/0x2cfcaff3289142e79173b856293d6128b6bd05c6}{\texttt{0x2cfc...05c6}}), and suspected CEX-DEX bots such as \href{https://arbiscan.io/address/0xf3ae7446b8a02550aD10A829Ad272D6fB7dBdDaB}{\texttt{0xf3ae...dDaB}}.\footnote{It is unlikely to be a DEX-DEX bot since the swap count equals the trade count.} The category also includes contracts from other DEXs and aggregators like Ramses, Camelot, PancakeSwap, SushiSwap, Odos, Dopex (now Stryke), 1inch, and many more.
To ensure that \textsc{cyclicArb} contracts were not incorrectly placed in the \textsc{other} category, we examined the transaction patterns of the identified contracts that were not labeled. We found that for all of them in the top 50\% of the gas consumers, $\geq 98\%$ of transactions that included at least one swap operation contained \textbf{exactly one swap}. This confirms that they are predominantly simple DEX interactions rather than complex cyclic arbitrage activities, justifying their classification.

\end{document}